\documentclass[authoryear,preprint,review,12pt]{elsarticle}

\usepackage{amssymb}
\usepackage{longtable}
\usepackage{amsmath}
\usepackage{acronym}
\usepackage{url}
\usepackage{xcolor}

\usepackage{hyperref}
\usepackage{cleveref}

\journal{Artificial Intelligence in Medicine}

\begin{document}

\begin{frontmatter}



\title{Deep Bayesian segmentation for colon polyps: Well-calibrated predictions in medical imaging}


\author[1]{Daniela L. Ramos \corref{cor1}}
\cortext[cor1]{\textit{email: dramosb@unbosque.edu.co}}
\author[1,2]{Hector J. Hortua}

\affiliation[1]{organization={Departmento de Matemáticas, Universidad El Bosque},
            city={Bogotá},
            country={Colombia.}}

\affiliation[2]{{Grupo Signos, Departamento de Matemáticas, Universidad El Bosque,},
city={Bogotá},
            country={Colombia.}}

\begin{abstract}
Colorectal polyps are generally benign alterations that, if not identified promptly and managed successfully, can progress to cancer and cause affectations on the colon mucosa, known as adenocarcinoma. Today advances in Deep Learning have demonstrated the ability to achieve significant performance in image classification and detection in medical diagnosis applications. Nevertheless, these models are prone to overfitting, and making decisions based only on point estimations may provide incorrect predictions. Thus, to obtain a more informed decision, we must consider point estimations along with their reliable uncertainty quantification. In this paper, we built different Bayesian neural network approaches based on the flexibility of posterior distribution to develop semantic segmentation of colorectal polyp images. We found that these models not only provide state-of-the-art performance on the segmentation of this medical dataset but also, yield accurate uncertainty estimates. We applied multiplicative normalized flows(MNF) and reparameterization trick on the UNET, FPN, and LINKNET architectures tested with multiple backbones in deterministic and Bayesian versions. We report that the FPN + EfficientnetB7 architecture with MNF  is the most promising option given its IOU of 0.94 and Expected Calibration Error (ECE) of 0.004, combined with its superiority in identifying difficult-to-detect colorectal polyps, which is effective in clinical areas where early detection prevents the development of colon cancer.
\end{abstract}




\end{frontmatter}



\section{Introduction}

Colorectal cancer is the second leading cause of cancer deaths
worldwide, both in terms of prevalence and mortality for both
genders. In 2020, it caused 935,000 deaths, accounting for $10\%$ of all cancer-related deaths. This highlights the importance of its study and early detection~\citep{WHO2020}. The 5-year survival rate for this type of cancer is around $65\%$ for all stages of the disease combined~\citep{seercolorectal}, but if detected early, this survival rate increases to $90\%$~\citep{cancerorg}. Colorectal polyps are known as direct precursors of this disease, if they are not treated adequately, effectively, and in time. The main tool to detect them is visually during the colonoscopy procedure, but this can lead to human errors during the diagnostic process, as studies have reported a rate of undetected polyps during the process, ranging from 6-28\%~\citep{Lee2017RiskFactors}.  Given the aforementioned reports, the importance of the development of automatic detection systems (ADS) for the accurate identification of colon polyps is evident. Recently, there have been several studies and approaches to automatic systems for polyps. One of the first proposals included morphology as WM-DOVA, where the authors implemented it in the CVC-CLINICDB dataset, being an approach to determine the presence and location of polyps, but it is not designed to accurately detect them at the pixel level~\citep{BJ2015WMDova}.  This was followed by an exploration of ADS based on convolutional neural networks at the semantic level using UNET-type~\citep{Tashk2023AutomaticSegmentation}, and FCN architectures~\citep{li_colorectal_2020}. These works reported notable results in terms of overall accuracy ($96\%$), but without the advantage of having uncertainties associated with the predictions. Finally, some works on segmentation using the CVC-CLINICDB database have reported the use of transformers like SegFormer \citep{wang2022stepwise}, Polyp-SAM \citep{li2023polypsam} and multiple CNN architectures, such as the double-UNET~\citep{jha2020doubleunet}, FCN-8 + VGG16, SegNet~\citep{WICKSTROM2020101619}, ResUNet++~\citep{jha2021comprehensive}, with acceptable results, but mostly report metrics such as $IOU < 90\%$ \citep{mei2024survey}.
On the other hand, quantification of uncertainties is a topic of great interest in Bayesian analysis. Bayesian methods offer probabilistic interpretations for predicted outcomes via a posterior distribution.  Although exact Bayesian inference with deep neural networks is computationally infeasible, the authors in~\citep{Gal2016} demonstrated that typical optimization of neural networks using dropout layers and L2 regularization can be seen as equivalent to performing Bayesian variational inference of a specific variational distribution \citep{KWON2020106816}.
In the field of medical image semantic segmentation, uncertainty estimation methods can be broadly classified
into Bayesian-based and Non-Bayesian-based methods, as reviewed by \citep{zou2023review}.  

Bayesian-based methods include several techniques for estimating uncertainty. These techniques involve using ensemble-based approaches that employ multiple models to capture different sources of variability. For example, MC dropout is a technique that introduces randomness by dropping out units from a neural network during inference, allowing for uncertainty estimation through repeated sampling. Other techniques are deterministic-based, aiming to develop algorithms that accurately estimate uncertainties \citep{wu2024colorectal}. Particularly for colon polyps,  the proposed method by~\citep{Gal2016}, which utilizes Monte Carlo estimator and dropout samples as seen in previous works like~\citep{WICKSTROM2020101619}, often produces inaccurate uncertainty estimates because deep neural networks trained with maximum likelihood estimation approaches do not provide precise confidence intervals. Although is not yet clear the cause of this miscalibration, \citep{pmlr-v70-guo17a} reported several experiments that present how the training and certain hyper-parameters impact the accuracy uncertainty estimates. The goal of this paper is to provide a road map to build accurate systems in terms of prediction performance and uncertainty estimates. We explore the use of different convolutional network structures with backbones and Bayesian approaches such as the multiplicative normalizing flows method and reparameterization trick to yield well-calibrated uncertainties.\\
The manuscript is structured as follows, in section \ref{sec:seccion2} to \ref{sec:seccion5}, we present models and methods, that include the introduction to concepts used to develop the work. Then, in section \ref{sec:seccion6}, we present experimental development and different architectures implemented. Next, in section \ref{sec:seccion7}, we present the main results and report the highest combination in terms of Bayesian approach and network architectures to predict polyps and their accurate uncertainties, Also, we employ feature importance methods to  understanding the correct interpretation of the model predictions. Finally, section \ref{sec:seccion8} presents conclusions along with appendix material.

\section{Bayesian Neural Networks} \label{sec:seccion2}
In the following, we introduce theoretical foundations of variational inference for Bayesian neural networks. It also covers measurement of model calibration, the association of uncertainties to predictions, and definition of recommended loss functions for binary segmentation cases.

\subsection{Variational Inference in Bayesian Neural Networks}

\label{sec:seccion2}
In the following, we introduce theoretical foundations of variational inference for Bayesian neural networks. It also covers measurement of model calibration, the association of uncertainties to predictions, and definition of recommended loss functions for binary segmentation cases.

\subsection{Variational Inference in Bayesian Neural Networks}

Within DNN framework, let $ \mathcal{D}=\lbrace(x_i,y_i)\rbrace_{i=1}^N$ where \textit{N} is  size of the sample and $x_i \in \mathcal{R}^d$ , $y_i=(y_i^{(1)},y_i^{(K)}) \in \lbrace(0,1)\rbrace^K $, \textit{d} is dimension of  input variables, \textit{K} is the number of different classes (\textit{output}), $\omega \in \Omega$ the vector of parameters for the network and $p(\omega)$ a prior on weights $\omega$. Posterior distribution is given by:

\begin{equation}
    p(\omega |\mathcal{D}) = \frac{p(\mathcal{D}|w)  \; p(\omega )}{p(\mathcal{D})} = \frac{ \prod_{i=1}^N p(y_i| x_i,\omega ) \; p(\omega )}{p(\mathcal{D})}
    \label{Eq.(1)}
\end{equation}

Predictive distribution (for a new pair $x_*,y_*$) is written as:

\begin{equation}
    p(y_*|x_*,\mathcal{D}) = \int_\omega p(y_*|x_*,\omega ) \;  p(\omega |\mathcal{D}) \; d\omega 
     \label{Eq.(2)}
\end{equation}

The computation of posterior $p(\omega|\mathcal{D})$ requires an integration over the entire lattice parameter space, which is computationally intractable. For this reason, variational inference methods with computation of the \textit{Kullback-Leibler} divergence are proposed:

\begin{equation}
KL\lbrace{q_\theta(\omega)\; || \; p(\omega|\mathcal{D})} \rbrace = \int_\Omega q_\theta(\omega) \log\frac{q_\theta(\omega)}{p(\omega|\mathcal{D})}
 \label{Eq.(3)}
\end{equation}

Hence, \textit{optimal distribution }is the distribution closest to the posterior among the pre-specified family $Q = {q_\theta(\omega) : \theta \in \Theta}$. For a \textit{mean-field approximation}, $Q$ is the family of fully factored gaussians, and \textit{i and j} are indices associated with the previous and current layer.

\begin{equation}
q_\theta(\omega) =\prod_{i=1}^L q_\theta(\omega_i) = \prod_{i,j} \mathcal{N}(w_{ij}; \mu_{ij},\sigma^2_{ij})
 \label{Eq.(4)}
\end{equation}

As divergence of KL is a measure of how similar two distributions are, minimizing this measure allows us to approximate the predicted distribution:

\begin{equation}
q_\theta(y_*|x_*) = \int_\Omega p(y_*|x_*,\omega ) q_\theta(\omega ) \; d\omega  \approx   p(y_*|x_*,\mathcal{D}) 
 \label{Eq.(5)}
\end{equation}

Solving the optimization problem by solving the minimum of the Kullback-Leiber divergence is equivalent to maximizing the \textit{evidence lower bound} (ELBO) \citep{Gal2016}, given by:

\begin{equation}
\mathcal{L}(\theta) = \int_\Omega q_\theta \log p(y|x,\omega)d\omega - KL(q_\theta(\omega )||p(\omega ))
 \label{Eq.(6)}
\end{equation}

Where $\mathcal{L}$ is a lower bound of 
\textit{log-likehood} of marginal posterior distribution. 

\subsection{Monte Carlo estimator}

Considering that the integration to compute the predicted distribution must be done over the entire $\Omega$ space, we consider a Monte Carlo estimator as follows:

\begin{equation}
\hat{p}_\theta(y_*|x_*)  = \frac{1}{T}
\sum_{i=1}^{T}p(y_*|x_*,w)q_\theta(w_t)
\end{equation}

Here $\lbrace w_t\rbrace_{t=1}^T$is a set of weight vectors randomly drawn from optimized variational distribution $q_\theta(w)$ with T number of samples. For a high value of T, it converges to the probability of $q_\theta(y_*|x_*)$ shown in Eq.(\ref{Eq.(5)}) for all $\omega \in \Omega$. \citep{KWON2020106816}

\subsection{Reparameterization Trick}

Part of the strategies for generating inference about posterior distribution and variance reduction is a sampling process during optimization, called reparameterization trick. Being $\omega$ the weights of the network, they can be written in terms of an auxiliary variable $\epsilon$:

\begin{equation}
 \omega \sim q(\omega|\theta) = g(\epsilon,\theta)
 \label{Eq.(7)}
\end{equation}

For $\epsilon \sim p(\epsilon)$ where  p is an independent distribution of  parameter $\theta$  that we want to optimize in  network training process. We get an estimation of  $q_\theta$ with:

\begin{equation}
\int_\Omega  q_\theta(\omega)  p(\omega) d\omega  \approx \frac{1}{K} \sum_{k=1}^K f(g(\epsilon,\theta)_K)
 \label{Eq.(8)}
\end{equation}

Let the distribution of weights $\omega \sim \mathcal{N}(\mu,\Sigma)$ we do a reparameterization using $\omega = \mu + \Sigma\epsilon$, where $\epsilon \sim \mathcal{N}(0,1)$. Using  Eq.\eqref{Eq.(8)}, we can approximate first term of  Evidence Lower Bound (ELBO) in Eq.\eqref{Eq.(6)}, this allows  estimation of sample gradient during training process by separating random part of sampling process from direct influence of the parameter being optimized. However, it is important to consider that one limitation of this method is that the weights of selected samples are the same for a given batch. This leads to a correlation of the gradients calculated in the samples. \citep{Hortua2020}.

\subsection{Multiplicative Normalizing Flows}
In the analysis of limit in Eq.\eqref{Eq.(6)}, ideal variational distribution is when $KL \lbrace q||p\rbrace $ equals zero. However, achieving this with \textit{mean field approximation} introduced in Eq.\eqref{Eq.(4)} is not feasible. For this purpose, we consider a more complex and flexible family of distributions that allows the true posterior distribution to be one of the possible solutions. By increasing the complexity, we expect significant performance enhancements because we can draw samples from a more reliable distribution that is closer to the true posterior. Multiplicative normalized flows (MNF), are a way to obtain  mentioned distributions through a combination of auxiliary random variables with normalization flows~\cite{Louizos2017}. By associating the parameter $\theta$ with a family of distributions to be compared over the posterior, and introducing an auxiliary latent variable in the form of a vector $z \sim q_\theta(z) \equiv q(z)^2$, the variational posterior can be represented mathematically as a blend of distributions

\begin{equation}
    q_\theta(w) = \int q_\theta(w|z)  q_\theta(z) dz
    \label{Eq.(10)}
\end{equation}

If the equation Eq.\ref{Eq.(4)} is rewritten including local reparametrizations, then posterior for fully connected layers will be \citep{garcíafarieta2023Bayesian}

\begin{equation}
    w \sim q(w|z) = \prod_{i,j} \mathcal{N}(w; z_i \mu_{ij},\sigma^2_{ij})
    \label{Eq.(11)}
\end{equation}

Let $f:\mathcal{R}^n \longrightarrow \mathcal{R}^n$, $f^{-1} = g$, and $g \circ f(z)=z$. A ramdom variable z with distribution $q(z)$ and $z'=f(z)$, satisfies

\begin{equation}
q(z') = q(z) \begin{vmatrix} det \; \frac{\partial f^{-1}}{\partial z'} \end{vmatrix} =  q(z) \begin{vmatrix} det \; \frac{\partial f}{\partial z} \end{vmatrix}^{-1}
\label{Eq.(12)}
\end{equation}

Then, having a composition $z_l = f_l(f_{l-1}(. . . f_1(z_0)))$, where $z_0 \sim q(z_0)$ are factorized gaussians like in Eq.\eqref{Eq.(4)}, for a sequence of $l$ invertible transformations, we have:

\begin{equation}
    \log q(z_l) = \log q(z_0) - \sum_{l=1}^L \log \begin{vmatrix} det \; \frac{\partial f_l}{\partial z_{l-1}} \end{vmatrix}
    \label{Eq.(13)}
\end{equation}

To calculate the posterior, implementing Bayes theorem $q(z_l)q(w|z_l)=q(w)q(z_l|w)$ and making use of an auxiliary distribution in the form $s(z_l|w,\phi)$ as in~\cite{Louizos2017}, with $\phi$ as parameter, we can get this auxiliary distribution as close as possible to this distribution with originals parameters $q(z_l|w)$, meaning KL divergence and its lower bound are given by:
\begin{align}
    -KL\lbrace q(w)|p(w) \rbrace \geq \mathbf{E}_{q(w,z_l)} [\ -KL[q(w|z_l)||p(w)] \nonumber  \\
  + \log q(z_l) + \log s(z_l|w,\phi)]\  \label{Eq.(14)}    
\end{align}

Initial term in right side can be determined analytically because its KL divergence calculated over two gaussians distributions. The second is determined by normalizing flow in Eq.\eqref{Eq.(13)} and given $z_0 = g_1^{-1}(g_{2}^{-1}(. . . g_l^{-1}(z_L)))$:

\begin{equation}
    \log s(z_l|w,\phi) = \log s(z_0|w,\phi) - \sum_{l=1}^L \log \begin{vmatrix} det \; \frac{\partial g^{-1}_l}{\partial z_{l}} \end{vmatrix}
    \label{Eq.(15)}
\end{equation}

By parameterizing the auxiliary posterior and transforming $g^{-1}_l$ into the form of a normalized flow~\cite{Louizos2017}, we obtain

\begin{equation}
    z_0 \sim s(z_l|w,\phi) = \prod_{i} \mathcal{N}(z_0;\tilde{\mu}_{i}(w,\phi),\tilde{\sigma}^2_{i}(w,\phi))
    \label{Eq.(16)}
\end{equation}

Here, we adopt parameterization of mean, represented as $\tilde{\mu}$, and variance, represented as $\tilde{\sigma}^2$, from \textit{masked real valued non volume preserving} (real NVP) like in \citep{dinh2017density} as 
option for normalizing flows.

\section{Observing calibration}\label{sec:seccion3}

A perfectly calibrated model is defined as one where prediction \( \hat{P} \) is a real probability in frequentist terms, i.e., it represents real probability that prediction is correct. This applies to a scenario with variables \( X \) and \( Y \), where \( X \in \mathcal{X} \), \( Y \) in \( \mathcal{Y} = \{0,1\} \). The joint distribution of \( X \) and \( Y \) is given by \( p(X, Y) = p(Y | X)p(X) \).  Otherwise, we have a neural network with input \( h(X) \) and prediction \( (\hat{Y}, \hat{P}) \), being \( \hat{Y} \) inference about the class and its associated probability \( \hat{P} \).Therefore, we have a calibrated model if~\citep{pmlr-v70-guo17a}

\begin{equation}
\mathbb{P} (\hat{Y} = Y | \hat{P} = p) = p, \forall p \in [0, 1]. 
\label{Eq.(17)}
\end{equation}

\subsection{Expected calibration error (ECE)}
Several metrics are available to measure a model calibration,  one of the most common and recognized is the so-called Expected Calibration Error (ECE). This metric, naturally derived from  Eq.(\ref{Eq.(17)}) represents the difference between prediction confidence and accuracy~\citep{wang2022calibrating}

\begin{equation}
ECE = \mathbf{E}_{\hat{p}} \left[ (\hat{Y} = Y | \hat{P} = p) - p \right],
\label{Eq.(18)}
\end{equation}

which is obtained by  computing the  weighted average of accuracy $acc(B_M)$ by partitioning p-space of predictions into M bins, where confidences are denoted by $conf(B_M)$, value $n$, and $|B_M|$ the number of pixels that fall into a bin. In semantic segmentation scheme \textit{n} represents the number of pixels

\begin{equation}
ECE = \frac{1}{n} \sum_{m=1}^{M} |B_M| |acc(B_M) - conf(B_M)|.
\label{Eq.(19)}
\end{equation}

A model is perfectly calibrated when its ECE is zero. The difference in each bin between accuracy and the confidences is represented visually by a \textit{gap} in the \textit{reliability diagrams}, a powerful tool for evaluating quality of  uncertainty estimations~\citep{wang2022calibrating}.

\subsubsection{ECE for semantic segmentation}

For ECE estimation, we adopt the approach followed in~\citep{wang2022calibrating}, where each pixel in an image is considered as a single sample, resulting in a total of NxNxI samples, where I is the total number of images to be evaluated and N is the size of the images. Then the ECE is calculated first in each image, and then later over all the images
\begin{equation}
ECE = \frac{1}{I} \sum_{i=1}^{I} ECE_i.
\label{Eq.(20)}
\end{equation}

\subsection{Reliability diagrams}\label{sec:seccion3.2}

Reliability diagrams are a visual representation of ECE, or equivalently, how well a model is calibrated. These graphs illustrate correlation between the expected accuracy of a sample and model confidences, using a partitioning of the prediction space into M bins. If model is perfectly calibrated, i.e. if the condition Eq.(\ref{Eq.(17)}) is satisfied  then, the relationship should be represented by an identity function. Any deviation from a perfect diagonal indicates a lack of calibration, implying that uncertainties are either under- or over-estimated~\citep{pmlr-v70-guo17a}. 

\section{Metrics and loss functions}\label{sec:seccion4}

\subsection{Metrics}
To evaluate the performance of the models, we considered IOU (Intersection over Union) since it measures the exact spatial similarity between areas segmented by the model and the masks. This metric, based on F-score, is particularly useful for evaluating accuracy of segmentation models in scenarios where high accuracy at edges of region of interest is critical~\citep{müller2022guideline}.

\begin{equation}
\text{IoU}_c = \frac{\sum_{i}{(y_i(c) \land \hat{y}_i(c))}}{\sum_{i}{(y_i(c) \lor \hat{y}_i(c))}},
\label{Eq.(21)}
\end{equation}
here, $c$ is the class, $y_i$ is mask value (ground truth) for class c,  $\hat{y}_i$ is prediction, $\land$ denotes \textit{and} operation, and $\lor$ denotes \textit{or} operation. As a supporting metric, \textit{recall} is also implemented, although this is less sensitive in isolation compared to F-score based metrics when assessing and comparing models. However, inclusion of recall helps us to provide a more comprehensive evaluation, allowing for a more nuanced understanding of a model performance and its ability to accurately identify ROI~\citep{müller2022guideline}  
\begin{equation}
\text{Recall} = \frac{TP}{TP + FN}.
\label{Eq.(22)}
\end{equation}

\subsection{Loss functions}\label{sec:seccion4.2}

The loss functions are crucial in training stage to produce accurate predictions, especially in semantic segmentation domain. Our work will employ  Jaccard loss, Dice loss, binary cross entropy, and total loss from  the python library \textit{Segmentation Models}~\citep{Yakubovskiy:2019}.

\subsubsection{Region based}\label{sec:4.2.1}
\vspace{2pt}
\begin{itemize}
\item Jaccard Loss: This loss function calculates intersection over union between region of interest (ROI) and region predicted by the model, to optimize the overlap between them

\begin{equation}
J(A, B) = 1 - \frac{{A \cap B}}{{A \cup B}},
\label{Eq.(23)}
\end{equation}
where \(A\) is region of interest in ground truth, and \(B\) is the region which is predicted by the model.

\item Dice Loss: Similar to Jaccard Loss, this loss function is also focused on calculating the intersection over union

\begin{equation}
D(A, B) = 1 - \frac{{2|A \cap B|}}{{|A| + |B|}}.
\label{Eq.(24)}
\end{equation}

This function is utilized to measure overlap or similarity between two sets and is commonly used in medical image segmentation tasks. The advantage of this loss function over Jaccard is that the overlap carries more weight in loss calculation, which is useful when proportion of pixels in one class, such as region of interest in an image, is significantly smaller than another. In other words, the goal is not only to maximize the proportion of overlapped region, but also to prioritize the exact level of overlap.\citep{losses}.
\end{itemize}

\subsubsection{Distribution based}
\vspace{5pt}

\begin{itemize}
\item Binary cross-entropy:  This is computed as  the difference between  actual distribution and the predicted distribution \citep{LossOdyssey}.

\begin{equation}
BCE(y, \hat{y}) = -\frac{1}{N}\sum_{i=1}^{N} y_i \log(\hat{y}_i) + (1-y_i) \log(1-\hat{y}_i),
\label{Eq.(25)}
\end{equation}
here, \(y\) is real value, \(\hat{y}\) is the prediction and \(N\) is number of samples.  
\item Total loss: This function takes into account both the similarity beetween regions of interest and the \textit{focus} on the minority class in cases of class imbalance in semantic segmentation task.
\begin{equation}
TL(y, \hat{y}) =  \mathcal{D}(A,B) + (0.5 * BFL).
\label{Eq.(63)}
\end{equation}
Dice loss is denoted by $\mathcal{D}$, and BFL is  \textit{Binary Focal Loss} function

\begin{equation}
BFL = -y \cdot \alpha \cdot (1-\hat{y})^\gamma \cdot \log(\hat{y}) \\ - (1-y) \cdot \alpha \cdot \hat{y}^\gamma \cdot \log(1-\hat{y}),
\label{Eq.(27)}
\end{equation}
where \(y\) is real value, \(\hat{y}\) is the prediction, \(\alpha\) is a weight and \(\gamma\) are a modulating parameter. The \textit{binary focal loss} function is an extension of cross entropy loss. It incorporates a gamma factor, known also as focusing parameter, which permits hard to classify pixels to have more severe penalties than those that are easier~\citep{Jadon_2020}.

\end{itemize}

\subsection{Neg-Log Likelihood}

The Negative Log Likelihood (NLL) is a function used to measure how closely a model fits the actual data. It is calculated based on number of samples $n$ and the distribution $p(Y|X)$

\begin{equation}
\text{NLL} = -\sum_{i=1}^{n} \log p((y_i|x_i)).
\label{Eq.(28)}
\end{equation}

In our case we will use it as a loss function for BNN models, any loss that includes an NLL is equivalent to minimizing the divergence \textit{Kullback-Leibler} in  Eq.\eqref{Eq.(3)},  or alternatively, it is a binary cross entropy computed between the distribution defined by training set and the probability distribution defined by model~\cite{goodfellow2016deep}.

\section{Dataset}\label{sec:seccion5}
\begin{figure}[!h]
    \centering
\includegraphics[width=0.31\textwidth,height=5.9cm]{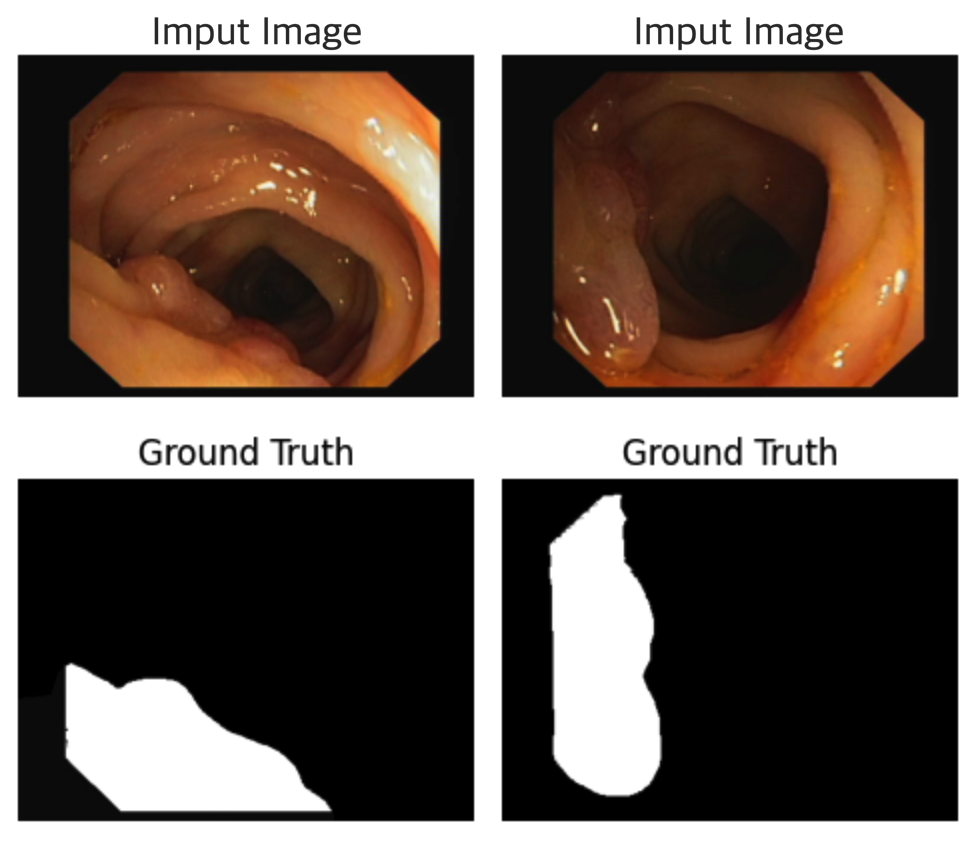}
     \caption{Example of some images in the dataset with their ground-truth~\citep{BJ2015WMDova}.}
     \label{fig:figura1}
\end{figure}

The CVC-CLINICDB database, which is a free and public database, will be used for this work. It was developed by~\citep{BJ2015WMDova}, and comprises 612 images extracted from colonoscopy videos and created for the study and development of automatic systems for detection and segmentation of colon polyps. Fig.(\ref{fig:figura1}) shows an preview of dataset. Images include ground truth and background (mucosa and lumen) and were obtained from 31 video sequences taken from 23 patients. The resolution of the images is $384 \times 288$.

\section{Experimental Setup}\label{sec:seccion6}

\subsection{Preprocessing}

Using the database referenced in Fig.(\ref{fig:figura1}), a binary segmentation task was conducted, class 0 represents the background and class 1 represents the polyp. The dataset was divided into three parts: training, validation, and test, with 70\%, 20\%, and 10\% respectively. All images were resized to 256x256 to eliminate black borders and facilitate network input. For the training images, preprocessing was performed in the following order
\begin{enumerate}
    \item  Adjust brightness, saturation, and contrast of the image randomly.
    \item  Randomly flip the image and mask to left or right.
    \item  Flip image and mask randomly up or down.
    \item  Normalization of pixels.
\end{enumerate}
The infrastructure put in place by Google Cloud Platform  uses an nvidia-tesla-t4 of 16 GB GDDR6 in an N1 machine series shared-core.
\subsection{Deterministic models}
At training phase, models were optimized using Adam optimizer with a batch size equals to eight. Early stopping was implemented by monitoring loss value on the validation set with a patience of 3. Four loss functions were used, as defined in Sec.(\ref{sec:seccion4.2}). The pipeline was built using Tensorflow v:2.15\footnote{https://www.tensorflow.org/} and Tensorflow-probability v:0.22\footnote{https://www.tensorflow.org/probability}. furthermore, we selected three architectures: Unet, Linknet, and FPN, using  python library \textit{Segmentation Models}\footnote{https://segmentation-models.readthedocs.io/en/latest/index.html}. This module offers several advantages, including ease of implementation, a choice of four model architectures have been proven to be effective for binary segmentation and 25 backbones with \textit{pre-trained} weights to achieve efficient convergence~\citep{Yakubovskiy:2019}. These architectures were tested with four loss functions mentioned in Sec.(\ref{sec:seccion4.2}) and three backbones that have been suggested for use in medical image segmentation: Seresnet101, Densenet169, EfficienNetB7~\citep{Abedalla2021}. A total of 36 iterations of deterministic models were performed for all possible combinations, as shown in the tables Tab.\ref{tab3}, Tab.\ref{tab4}, Tab.\ref{tab5}.

\subsection{Bayesian models}
\vspace{5pt}

\subsubsection{Multiplicative Normalizing Flows (MNF)}
\vspace{2pt}

\begin{figure}[!h]
\centering
\includegraphics[scale=.43]{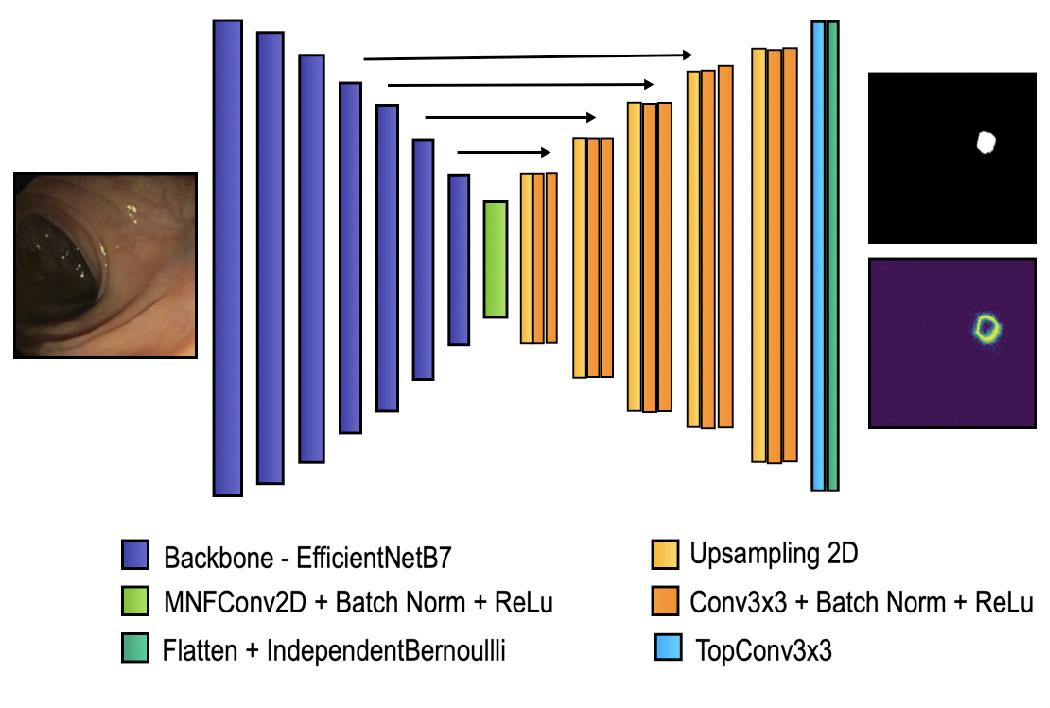}
\caption{Unet architecture: MNF layer is positioned at the output of the backbone.}

\label{fig:figura3}
\end{figure}

\begin{figure}[!h]
\centering
\includegraphics[scale=.42]{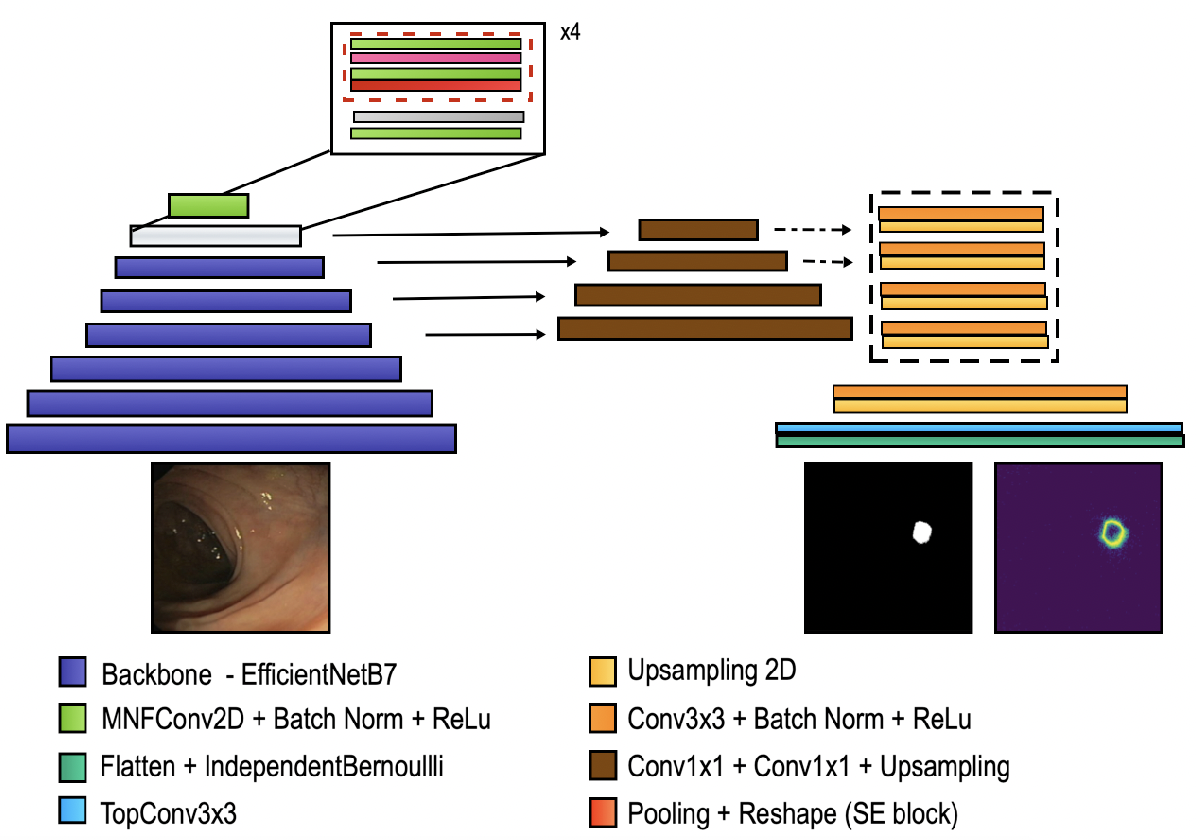}
\caption{FPN architecture: MNF layers are placed in all layers of last block in the backbone.} 

\label{fig:figura4}
\end{figure}

\begin{figure}[!h]
\centering
\includegraphics[scale=.42]{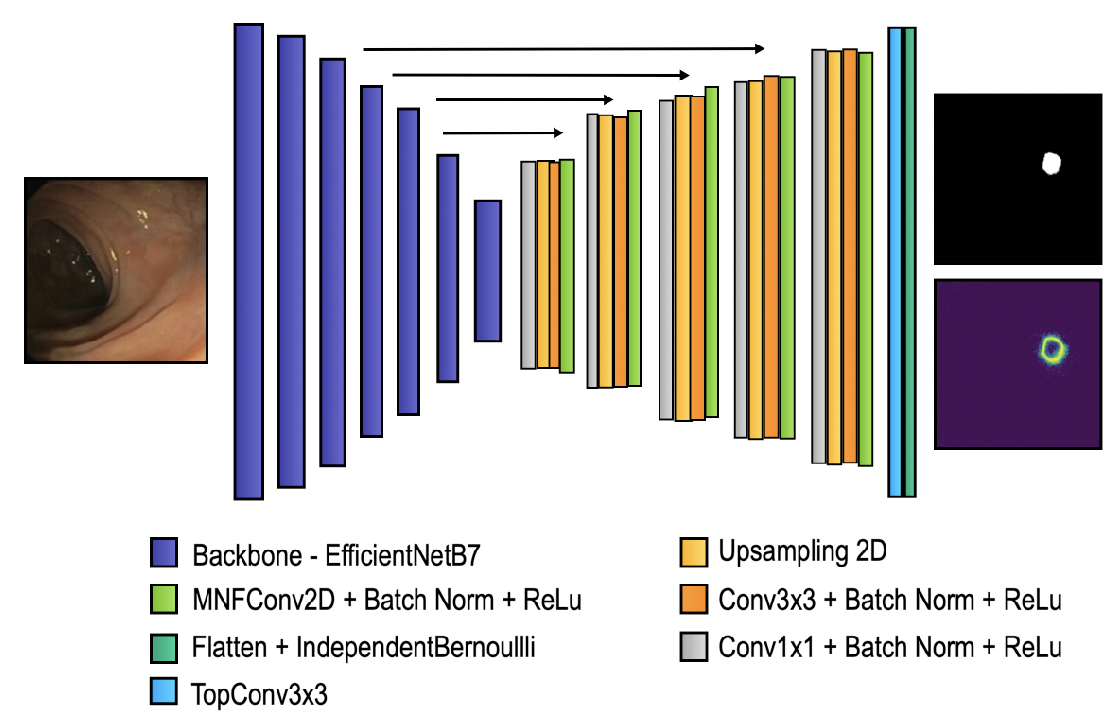}
\caption{Linknet architecture: MNF layers are placed in last layer of each decoder block.}

\label{fig:figura5}
\end{figure}

We adapted deterministic architectures to a Bayesian approach using the module \textit{models} from  \textit{segmentation models}. To carry out this task, we utilized  MNFConv2D class from tf-mnf module\footnote{https://github.com/janosh/tf-mnf/tree/main}, replacing some strategic Conv2D Tensorflow layers in this code. Moreover, we have modified output layer of these architectures by adding a layer \textit{Independent Bernoulli} from Tensorflow-probability module. Three architectures with highest IOU metric in test~\ref{appendix:Ap.C},  Unet + EB7, FPN + EB7, and Linknet + EB7, were evaluated with three different configurations each, resulting in a total of nine models.  The MNFConv2D layers were strategically placed in the networks. All models were trained using the defined NLL loss Eq.\eqref{Eq.(28)} function. \\
The nine modified configurations are as follows:
\begin{enumerate}
\item UNET: Backbone output - Fig.(\ref{fig:figura3}), all layers of the final block of the backbone, last layer of each decoder.
\item FPN: Backbone output,  all layers of the final block of the backbone - Fig.(\ref{fig:figura4}), output concatenate + output last pyramidal block.
\item Linknet: Backbone output, all layers of the final block of the backbone, last layer of each decoder Fig.(\ref{fig:figura5}).
\end{enumerate}

\subsubsection{Reparametrización Trick}

Considering the best combination of backbone and layer location for each architecture mentioned in the previous section, we replaced light green layers shown in Fig.(\ref{fig:figura3}), Fig.(\ref{fig:figura4}) and Fig.(\ref{fig:figura5}) with \textit{Conv2DReparameterization} layers from the Tensorflow-probability library. Results of iterations can be found in Tab.(\ref{tab1}).

\section{Results}\label{sec:seccion7}

Considering results achieved in all architectures Unet, FPN and Linknet, and using IOU as the main metric, and secondly, recall, it was determined that EfficientNetB7 is best backbone in terms of performance in iterations of Tab.\ref{tab3}, Tab.\ref{tab4}, Tab.\ref{tab5}. In particular, binary cross entropy loss function was found to be the most efficient for Unet and Linknet architectures. Conversely, for FPN, total loss function was the best alternative. The top-performing model in iterations was Linknet+EB7+BinaryCE, achieving an IOU of 0.941 in test. Otherwise, the model with worst performance was FPN+ Densenet169+ Total loss, with an IOU of 0.78 and recall 0.72 in test. Upon analyzing the tables in \ref{appendix:Ap.C}, it is found that the best configuration for all iterations performed with Densenet169 backbone was FPN - BinaryCE, with $IOU=0.92$. 
The results of the combinations performed with different architectures and loss functions in Densenet169 show the presence of many false negatives. This is evidenced by the recall, which is consistently below 0.8 in most combinations and lower on average than other iterated backbones. In particular, the combination Densenet169 + Total Loss does not work well in any of the architectures and therefore, is not recommended. Similarly, for Seresnet101, the best model was the combination given by FPN - BinaryCE, with an IOU of $0.92$. During the deterministic iterations with this backbone, we observed that the IOU was higher than $0.82$ in all iterations. A lower sensitivity to detection is seen when employing the Linknet architecture with this backbone, this is evidenced by a higher number of false negatives compared to Unet and FPN. Therefore, the use of the Linknet + Seresnet101 configuration is not recommended either. Furthermore, we could observe that the region-based loss functions Sec.(\ref{sec:4.2.1}) did not provide a real benefit in improving ROI detection in contrast to other loss functions. This might imply that the class imbalance between the polyp and the background would not be significantly affecting the performance of the models. Moreover, in iterations performed by introducing MNF layers Tab.(\ref{tab2}),  it was found that the best configurations are those in Fig.(\ref{fig:figura3}), Fig.(\ref{fig:figura4}) y Fig.(\ref{fig:figura5}). 
MNF model that performed the best was Linknet in Fig.(\ref{fig:figura5}) configuration, achieving an IOU of 0.94 in test.

\begin{table}[!h]
\caption{\label{tab2} Results in test dataset for models implementing MNFConv2D layers.}
\vspace{4pt}
  \scalebox{0.72}{
\begin{tabular}{ccllll}
\hline
\begin{tabular}[c]{@{}c@{}}Models\\ MNF layers\end{tabular} &
  \begin{tabular}[c]{@{}c@{}}Layers\\ position\end{tabular} &
  IOU &
  Recall &
  \begin{tabular}[c]{@{}l@{}}False \\ negatives\end{tabular} &
  \begin{tabular}[c]{@{}l@{}}False \\ positives\end{tabular} \\ \hline
\begin{tabular}[c]{@{}c@{}}UNET           \\ EfficienNetB7\end{tabular} &
  \begin{tabular}[c]{@{}c@{}}Backbone output\\ \\ \\ All layers of final block\\ in backbone\\ \\ Last layer of each decoder\\ block \end{tabular} &
  \begin{tabular}[c]{@{}l@{}}\textbf{0.9319}\\ \\ \\ 0.919\\ \\ \\ 0.9302\end{tabular} &
  \begin{tabular}[c]{@{}l@{}}\textbf{0.94}\\ \\ \\ 0.879\\ \\ \\ 0.898\end{tabular} &
  \begin{tabular}[c]{@{}l@{}}40680\\ \\ \\ 81675\\ \\ \\ 68573\end{tabular} &
  \begin{tabular}[c]{@{}l@{}}43208\\ \\ \\ 13959\\ \\ \\ 14482\end{tabular} \\ \hline
\begin{tabular}[c]{@{}c@{}}FPN        \\ EfficienNetB7\end{tabular} &
  \begin{tabular}[c]{@{}c@{}}Backbone output\\ \\ \\ All layers of final block\\ in backbone\\ \\ Final stage: output\\ concatenate + output\\ last pyramidal block \end{tabular} &
  \begin{tabular}[c]{@{}l@{}}0.892\\ \\ \\ \textbf{0.937}\\ \\ \\ \\ 0.926\end{tabular} &
  \begin{tabular}[c]{@{}l@{}}0.829\\ \\ \\ \textbf{0.925}\\ \\ \\ \\ 0.894\end{tabular} &
  \begin{tabular}[c]{@{}l@{}}115634\\ \\ \\ 40515\\ \\ \\ \\ 71404\end{tabular} &
  \begin{tabular}[c]{@{}l@{}}11549\\ \\ \\ 21424\\ \\ \\ \\ 16856\end{tabular} \\ \hline
\begin{tabular}[c]{@{}c@{}}LINKNET        \\ EfficienNetB7\end{tabular} &
  \begin{tabular}[c]{@{}c@{}}Backbone output\\ \\ \\ All layers of final block\\ in backbone\\ \\ Last layer of each decoder\\ block\end{tabular} &
  \begin{tabular}[c]{@{}l@{}}0.9326\\ \\ \\ 0.936\\ \\ \\ \textbf{0.9402}\end{tabular} &
  \begin{tabular}[c]{@{}l@{}}0.911\\ \\ \\ 0.909\\ \\ \\ \textbf{0.921}\end{tabular} &
  \begin{tabular}[c]{@{}l@{}}60073\\ \\ \\ 61704\\ \\ \\ 53795\end{tabular} &
  \begin{tabular}[c]{@{}l@{}}20758\\ \\ \\ 13786\\ \\ \\ 17797\end{tabular} \\ \hline
\end{tabular}}
\end{table}
Fig.(\ref{fig:figura4}) shows the performance generated by a FPN architecture with an IOU of 0.937. Notice that, it yields  higher recall, 0.925, compared to Linknet's 0.92 value. For iteration in which Bayesian MNF layers were replaced with reparameterization layers, we found that Unet model performed the most successfully, with an $IOU=0.92$ in test.
\subsubsection{Transformers: Segformer}

Recent studies have shown that transformer-based architectures are effective for semantic segmentation, so it is important to consider the potential benefits of using SegFormers in this work. SegFormers use multi-scale overlapping windows and a hybrid attention mechanism to optimize both global and local features~\citep{xie2021segformer,wu2024colorectal}. This could improve the models ability to detect subtle variations in polyps characteristics.  Based on the above, we conducted iterations with SegFormer B0 and SegFormer B5 architectures. However, the latters showed that the SegFormer models were not performing well. The primary metrics, IOU and recall, yielded results of less than 0.7, leading us to discontinue further iterations with these models for this case.

\subsection{Architecture}
\subsubsection{Unet}

The initial iterations for UNET produced results that are summarized in Tab.(\ref{tab3}). The best results were obtained for the EfficientNetB7 backbone, with an IOU in test greater than 0.9. For the iterations performed with EB7 with four loss functions Sec.(\ref{sec:seccion4.2}), binary cross entropy performed better in all metrics evaluated, in comparison to other functions. Focusing on iterations with Seresnet101 and Densenet169, it is clear that both models show a acceptable overall performance, with IOU results consistently, above 0.8. Tab.(\ref{tab3}). In this case, model Unet + Densenet + Total Loss  exhibits the lowest recall, of 0.69. Conversely, Unet + Seresnet + BCE model achieved a higher recall (0.89). Therefore, it can be inferred that the latter offers a more balanced performance.

 \begin{table}[!h]
\caption{\label{tab1} Results in test dataset for models implementing reparameterization trick (RT) }
\vspace{5pt}
\scalebox{0.78}{
\begin{tabular}{ccllll}
\hline
\begin{tabular}[c]{@{}c@{}}Models\end{tabular} &
  \begin{tabular}[c]{@{}c@{}}Layers\\ position\end{tabular} &
  IOU &
  Recall &
  \begin{tabular}[c]{@{}l@{}}False \\ negatives\end{tabular} &
  \begin{tabular}[c]{@{}l@{}}False \\ positives\end{tabular} \\ \hline
\begin{tabular}[c]{@{}c@{}}UNET           \\ EfficienNetB7\end{tabular} &
  \begin{tabular}[c]{@{}c@{}}Backbone \\ output\end{tabular} &
  \textbf{0.921} &
  0.891 &
  58513 &
  19597 \\ \hline
\begin{tabular}[c]{@{}c@{}}FPN        \\ EfficienNetB7\end{tabular} &
  \begin{tabular}[c]{@{}c@{}}All layers of final \\ block in \\ backbone\end{tabular} &
  0.908 &
  0.885 &
  61792 &
  30662 \\ \hline
\begin{tabular}[c]{@{}c@{}}LINKNET        \\ EfficienNetB7\end{tabular} &
  \begin{tabular}[c]{@{}c@{}}Last layer of \\ each decoder\\ block\end{tabular} &
  0.906 &
  \textbf{0.946} &
  28736 &
  71331 \\ \hline
\end{tabular}}
\end{table}

Regarding results of Bayesian iterations, we made a direct comparison between UNET + EB7 deterministic architecture Tab.(\ref{tab3}) and the one with MNF layers Tab.(\ref{tab2}). IOU metric and accuracy in test dataset remained unchanged, while recall increased from 0.9 to 0.94. However, accuracy decreased from 0.97 to 0.93, indicating that MNF model is more sensitive to regions classified as polyps, resulting in a higher number of false positives. If we contrast this result with implementation of reparameterization layers in this structure Tab.(\ref{tab1}), the metrics decreased, particularly recall (from 0.94 to 0.89) and IOU (from 0.93 to 0.92), resulting in increased false negatives.

\subsubsection{Linknet}

In deterministic Linknet iterations, the best result was achieved again with EB7. In regard to loss functions, binary cross entropy outperformed the other functions in all evaluated metrics. Based on the analysis, it can be concluded that both Linknet combined with Seresnet101 and Densenet169 produce acceptable IOU results, with a score above 0.8 Tab.(\ref{tab5}). However, they show lower recall than EB7 and then, a higher number of false negatives. Despite decent IOU performances, these configurations would not be optimal as they might have issues with under-detection. For Linknet + EB7 contrasting deterministic with MNF method, test metrics remain unchanged, except for a slight increase in precision from 0.96 to 0.97, resulting in a decrease in false positives.  In case of reparameterization, performance decreases, lowering IOU from 0.94 to 0.9 and precision, from 0.96 to 0.87. Recall enhances from 0.92 to 0.95, generating high sensitivity and the number of false positives. 

\subsubsection{FPN}
For FPN architecture in deterministic case, the best result was achieved with EB7. Its IOU and Recall in test were slightly higher than the others, with $IOU > 0.9$. Among loss functions, total loss was superior to others, achieving an IOU of 0.93 and a recall of 0.94. In contrast to others architectures, EB7+BCE was worst loss function with a $IOU = 0.89$ and $recall = 0.86$. Analysis of Tab.(\ref{tab4}) indicates Densenet169 with FPN has an IOU of approximately 0.8, except for Binary CE where it performed well with an IOU and recall of 0.92. However, this combination has lower recall overall and increased false negatives, particularly when using Densenet169 with FPN and Total loss. On the other hand, Seresnet101 with FPN has similar IOU results at 0.9, except for Jaccard loss, which had an IOU of 0.82 and a significant increase in false negatives. Despite acceptable IOU performance, these combinations exhibit low recall, which may result in under-detection issues. In the Bayesian FPN+EB7 counterpart, IOU slightly enhanced from 0.93 to 0.937, while precision improved from 0.94 to 0.96. At the same time, recall decreased from 0.94 to 0.925, reducing the number of false positives and improving performance. When comparing results obtained through reparameterization trick, IOU drops from 0.93 to 0.91 and recall drops to 0.88, thereby increasing the number of pixels with false negatives.

\subsection{Reliable analysis}
\begin{figure}[!h]
\centering
\includegraphics[scale=.46]{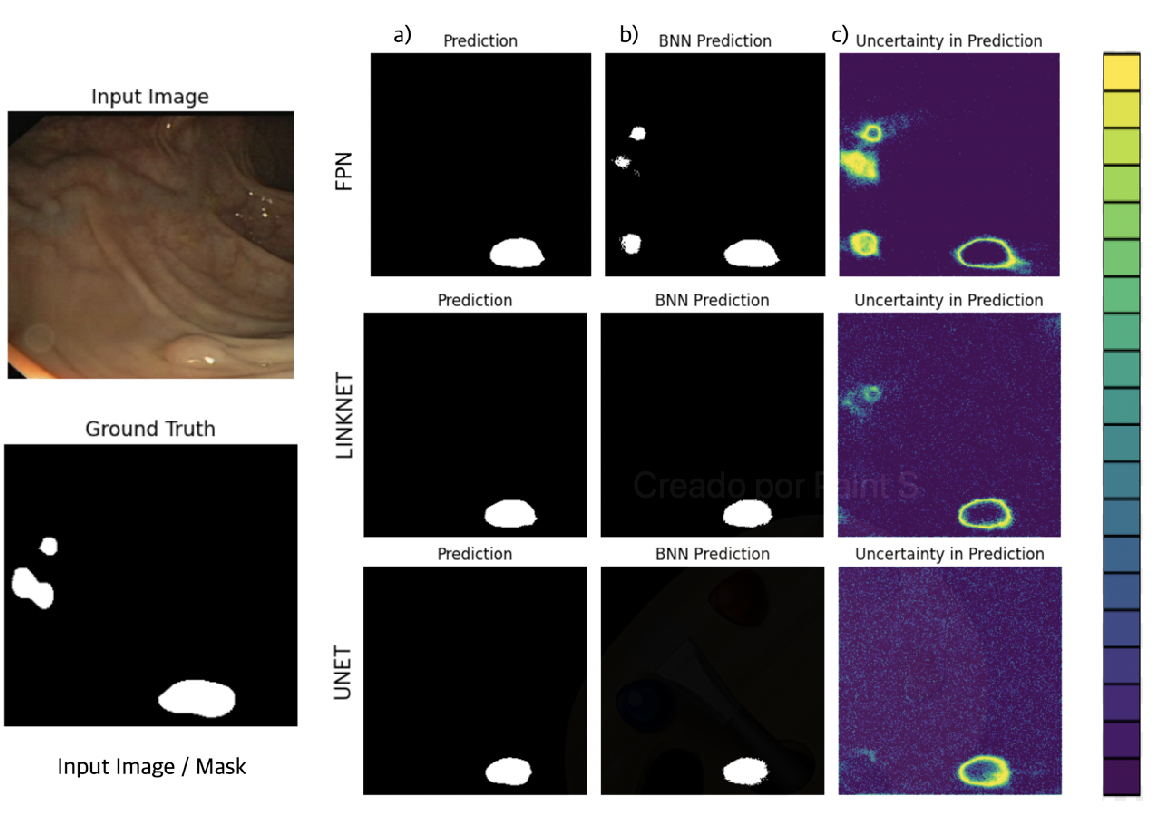}
\caption{(a) Deterministic prediction, (b) BNN prediction with MNF layers, and (c)uncertainty maps for the same input image employed UNET, LINKNET and FPN architectures.
}
\label{fig:figura6}
\end{figure}

Fig.(\ref{fig:figura6}) illustrates a comparison between BNN models with MNF layers Fig.(\ref{fig:figura6}b) and their respective deterministic versions (Linknet+EB7+BinaryCE, UNET+EB7+BinaryCE, FPN+EB7+Total Loss), Fig.(\ref{fig:figura6}a), can be appreciated. In all three cases, deterministic versions were unable to accurately detect smaller polyps present in the example image. To calculate the mask for BNN models, we take 50 predictions over the input image, average them, and then binarize the result.

Concerning heat maps, we can see that models have a low uncertainty in their predictions, except at the edges of the polyps and in those cases where they are difficult to detect. In the example provided, FPN model is the most sensitive, particularly to small polyps and image reflections compared to other models. In contrast, Linknet exhibited a more balanced performance and showed moderate sensitivity to these challenging cases. On the other hand, UNET model did not detect the presence of small polyps. It is evident from the estimation of the final prediction in Fig.(\ref{fig:figura6}b) that Linknet and UNET did not report the polyps present in the ground truth, while FPN was able to detect them.  These results demonstrate the significance of evaluating and comparing different capabilities in polyp detection, also, it is important to consider not only performance metrics but also the visual uncertainty presented by each model. Moreover, a visual representation of deterministic predictions, BNN predictions, and corresponding uncertainties for Bayesian networks with reparameterization trick can be found in Fig.(\ref{fig:figura8}). It shows that all models exhibit high uncertainty in the edge regions and circular details in the image, nevertheless, Linknet model, seems to be more sensitive to these uncertainties, as reflected in the visual representations. Otherwise, FPN model shows a more stable heat map and less uncertainty. It is worth mentioning that none of the three models succeeds in capturing small polyps present in ground truth for three masks in column b), which represent the mean of BNN predictions. Besides, only linknet model is observed to have high uncertainty in this particular region in uncertainty maps.

\begin{figure}[!h]
\centering
\includegraphics[scale=.36]{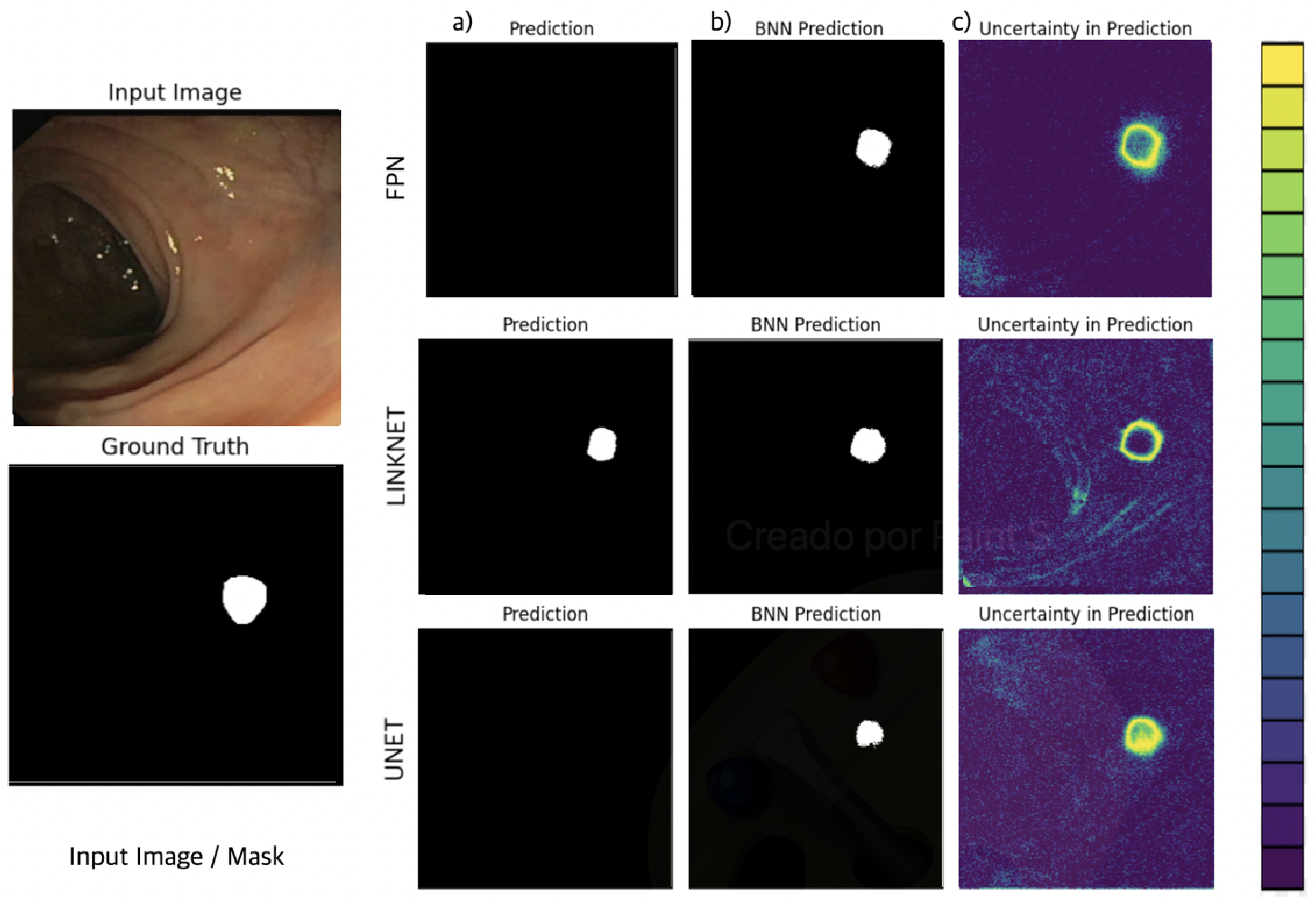}
\caption{(a) Deterministic prediction, (b) BNN prediction with RT layers, and (c) Uncertainty maps for the same input image employed UNET, LINKNET and FPN architectures.}
\label{fig:figura6a}
\end{figure}

The performance in the case of non-easily visible polyps, such as the input image in Fig.(\ref{fig:figura6a}), underscores the advantage of employing Bayesian neural networks over deterministic architectures. By examining column Fig.(\ref{fig:figura6a}a), it can be seen that while the deterministic Linknet model was able to detect the polyp, the deterministic FPN and Unet models failed to identify the afflicted region. As demonstrated in columns Fig.(\ref{fig:figura6a}b) and c), the Bayesian predictions with the reparameterization trick and the uncertainty maps were able to successfully identify the anomaly. This highlights the importance of uncertainties, particularly in this field where early detection is high priority.

\subsection{Model calibration}
\begin{figure}[t]
\centering
\includegraphics[scale=.49]{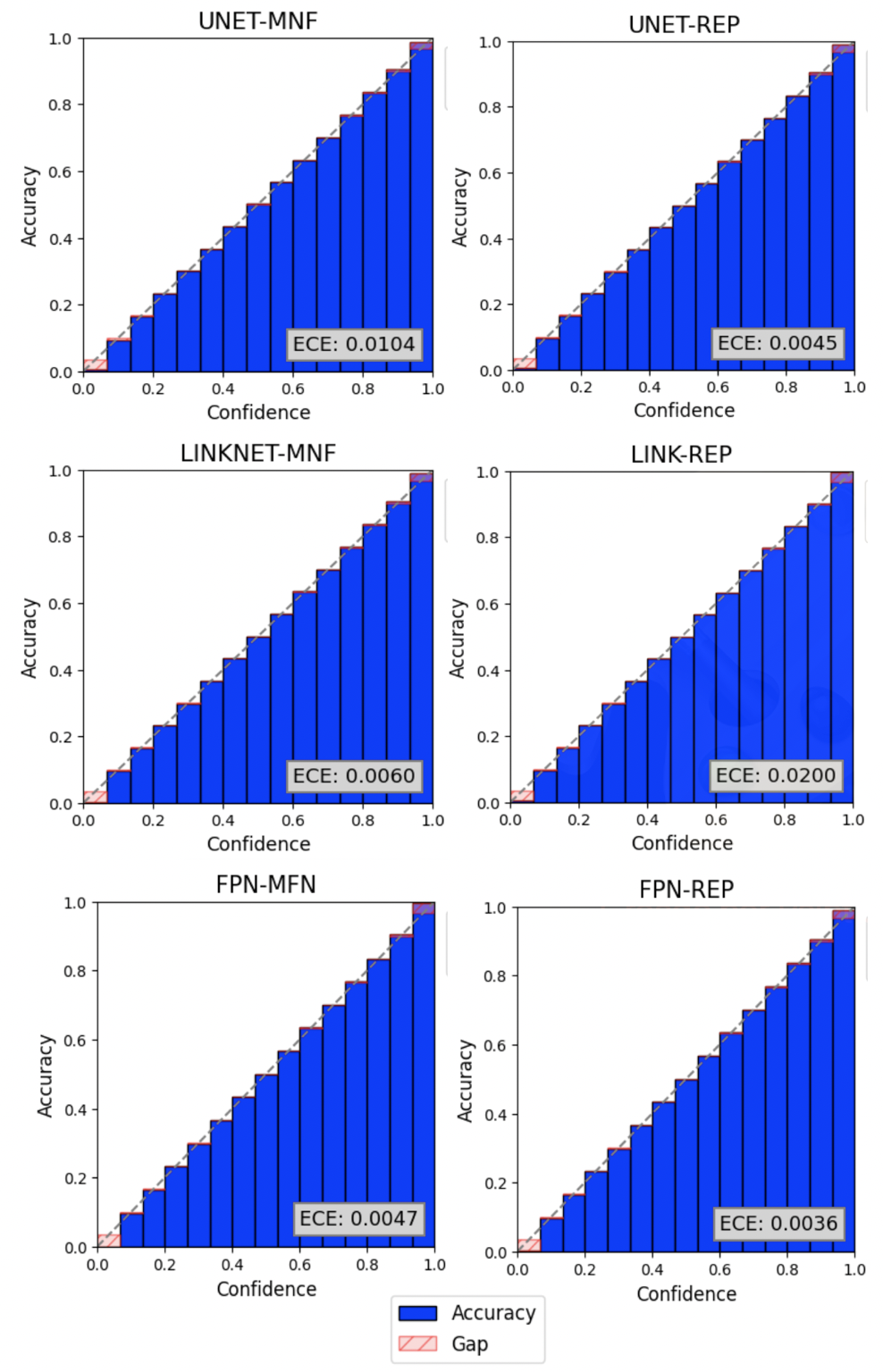}
\caption{Reliability diagrams with ECE metric using M=15 bins, for models with MNF layers vs. models with reparameterization layers. A smaller gap and an ECE close to zero indicate a better calibration.}
\label{fig:figura7}
\end{figure}

\vspace{2pt}
In this section, we develop a detailed analysis of reliability diagrams for each of the six models in Sec.(\ref{sec:seccion3.2}). Reliability diagrams provide a visual representation of the predictions and uncertainties predicted by models. The graphs used in this study were adapted from~\citep{wang2022calibrating}, with specific modifications for the semantic segmentation task. In this case, each pixel of an image was treated as an individual sample, resulting in a total of  256x256xM samples, being M the total number of images in test set. These samples were used to calculate plots shown in Fig.(\ref{fig:figura7}) and to report the Expected Calibration Error (ECE) using Eq.(\ref{Eq.(19)}) and Eq.(\ref{Eq.(20)}) with M=15 bins. ECE value can be found in the bottom corner of each plot. Fig.(\ref{fig:figura7}) illustrates how the models are well-calibrated, being models with best calibration those based on FPN, with $ECE = 0.004$, while model with lowest calibration was Linknet with reparameterization layer, showing an ECE one order higher $ECE = 0.02$. This can be related to what discussed over Fig.(\ref{fig:figura6}c) and Fig.(\ref{fig:figura8}c), where uncertainty maps of FPN models have a higher contrast in the palette used, since background color is uniform compared to what was observed in other architectures. For UNET architectures, when evaluating diagrams of the versions with MNF approach and reparameterization layers, it is obtained that both models are well calibrated, being better the version with RT, since its ECE is lower 0.01 against 0.0045. This can be related to the increase of precision value in test set, changing from 0.93 Tab.(\ref{tab2}) to 0.96 value reported in Tab.(\ref{tab1}), decreasing then the number of false positives. For Linknet, it is found that implementation of MNF layers has a lower ECE of 0.006 versus 0.02 of reparameterized trick version, indicating a higher reliability in probabilities and uncertainties predicted by first model. This calibration advantage is directly related to better prediction quality and capability. On the other hand, reparameterized trick version has a higher rate of false positives, with a natural decrease in its accuracy value. Moreover, for FPN, a similar behavior is observed for both versions, with an ECE of 0.0047 against 0.0036 for version with reparameterization trick, indicating that the latter is slightly better. This is also evidenced in metrics, where we can observe a minor improvement in accuracy, going from 0.93 to 0.95 (Tab.(\ref{tab1}),Tab.\ref{tab2}). This not only shows a improved performance, but also implies a reduction of false positives.

\subsection{Feature Importance}
\begin{figure}[h!]
\centering
\includegraphics[scale=.27]{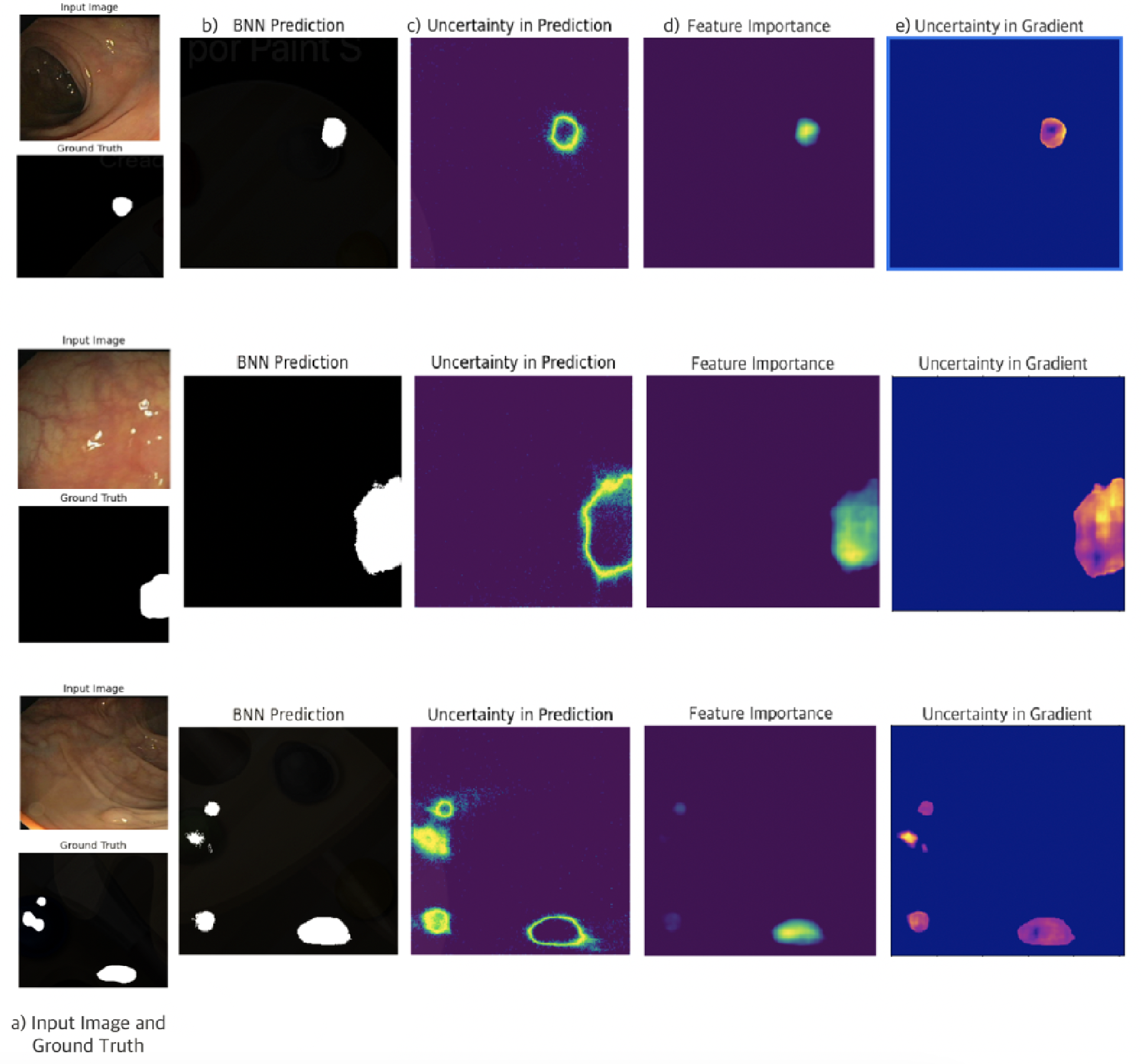}
\caption{(a) Input image and ground truth, (b) BNN prediction with FPN model with MNF layers, (c) Uncertainty maps d) Feature importance and e) Gradient uncertainty }
\label{fig:figura8fi}
\end{figure}

Following the methodology in~\cite{WICKSTROM2020101619}, We compute the feature importance of the segmentation images via image-gradient approach~\cite{simonyan2014deep} to interpret the results generated by the networks.  Fig.~\ref{fig:figura8fi} illustrates the crucial pixels for the segmentation process, particularly in areas containing polyps.
 Most notable features are observed near the borders of polyps. Bayesian techniques are not affected by changes far from the regions of interest, demonstrating robustness and interpretability. However, the influence zone surpasses the polyps borders, suggesting that the prediction also takes into account the global setting. Deterministic networks are inadequate in detecting atypical regions in situations with extremely small polyps, resulting in unsatisfactory segmentation outcomes. This is supported by the feature visualization shown at the top of Fig.~\ref{fig:figura6a}. In contrast, uncertainty estimates can identify areas where polyps may be present and offer crucial insights into the unreliability of neural networks in making predictions in certain pixel locations.  
\section{Conclusions}\label{sec:seccion8}

The result of this study shows that Bayesian models evaluated stand out for their good performance, since they have an IOU in test set consistently above 0.9, which shows efficiency of  architectures tested for semantic segmentation of medical images. The architecture based on Linknet + EfficientnetB7 demonstrated good results in both, deterministic and its Bayesian configuration (MNF layers). It presented a good calibration as well as a balanced option in visual terms and with adequate sensitivity for detecting colorectal polyps. However, FPN architectures with Bayesian layers are noteworthy for their ability to detect polyps that are difficult to identify with naked eye. They performed better than other architectures due to better calibration and uncertainty maps with more contrast between background and polyp edge. According to this study, FPN+ EfficientnetB7 with MNF or reparameterization trick layers was found to be the most suitable option for this aspect. Linknet configuration is also considered a viable option, but caution should be employed in scenarios involving smaller or difficult-to-visualize polyps. Finally, Unet version with reparameterization layers outperformed its MNF counterpart by better handling false positives, resulting in a tighter calibration. However, reparameterization trick approach in Linknet showed lower performance in terms of calibration, leading to an increase in false positives and overestimation of uncertainties in heat maps. This configuration is less recommendable compared to the ones studied, particularly for clinical case of polyp detection where accuracy is critical.
The scripts used for different experiments shown in this paper can be found in  \href{https://github.com/JavierOrjuela/medical-interpretability-polyp-detection}{medical-interpretability-polyp-detection
}.
\section*{Acknowledgements}
This paper is based on work supported by the Google Cloud
Research Credits program with the award GCP19980904. HJH acknowledges
support from the grant provided by the Google Cloud Research Credits program.

\section*{Abbreviations list}

The following is a list of abbreviations used throughout the document.

\begin{acronym}
\acro{ACS}[ACS]{American Cancer Society}
\acro{ADS}[ADS]{Automatic Detection Systems}
\acro{BNN}[BNN]{Bayesian Neural Networks}
\acro{D169}[D169]{DenseNet169}
\acro{DNN}[DNN]{Deep Neural Networks}
\acro{EB7}[EB7]{EfficientNetB7}
\acro{ECE}[ECE]{Expected Calibration Error}
\acro{MNF}[MNF]{Multiplicative Normalizing Flows}
\acro{MIS}[MIS]{Medical Image Segmentation}
\acro{NCI}[NCI]{National Cancer Institute}
\acro{NLL}[NLL]{Negative Log-Likelihood}
\acro{RT}[RT]{Reparameterization Trick}
\acro{S101}[S101]{SeresNet101}
\acro{WHO}[WHO]{World Health Organization}
\end{acronym}

\bibliographystyle{model2-names.bst}
\biboptions{authoryear}
\bibliography{refs}

\begin{thebibliography}{32}
\expandafter\ifx\csname natexlab\endcsname\relax\def\natexlab#1{#1}\fi
\providecommand{\url}[1]{\texttt{#1}}
\providecommand{\href}[2]{#2}
\providecommand{\path}[1]{#1}
\providecommand{\DOIprefix}{doi:}
\providecommand{\ArXivprefix}{arXiv:}
\providecommand{\URLprefix}{URL: }
\providecommand{\Pubmedprefix}{pmid:}
\providecommand{\doi}[1]{\href{http://dx.doi.org/#1}{\path{#1}}}
\providecommand{\Pubmed}[1]{\href{pmid:#1}{\path{#1}}}
\providecommand{\bibinfo}[2]{#2}
\ifx\xfnm\relax \def\xfnm[#1]{\unskip,\space#1}\fi
\bibitem[{Abedalla et~al.(2021)Abedalla, Abdullah, Al-Ayyoub and Benkhelifa}]{Abedalla2021}
\bibinfo{author}{Abedalla, A.}, \bibinfo{author}{Abdullah, M.}, \bibinfo{author}{Al-Ayyoub, M.}, \bibinfo{author}{Benkhelifa, E.}, \bibinfo{year}{2021}.
\newblock \bibinfo{title}{Chest x-ray pneumothorax segmentation using u-net with efficientnet and resnet architectures}.
\newblock \bibinfo{journal}{PeerJ Computer Science} \bibinfo{volume}{7}, \bibinfo{pages}{607}.
\newblock \DOIprefix\doi{10.7717/peerj-cs.607}.
\bibitem[{ACS(2022)}]{cancerorg}
\bibinfo{author}{ACS}, \bibinfo{year}{2022}.
\newblock \bibinfo{title}{Detección temprana del cáncer colorrectal}.
\newblock \URLprefix \url{https://www.cancer.org/es/cancer/tipos/cancer-de-colon-o-recto/deteccion-diagnostico-clasificacion-por-etapas/deteccion.html}.
\bibitem[{Azad et~al.(2023)Azad, Heidary, Yilmaz, Hüttemann, Karimijafarbigloo, Wu, Schmeink and Merhof}]{losses}
\bibinfo{author}{Azad, R.}, \bibinfo{author}{Heidary, M.}, \bibinfo{author}{Yilmaz, K.}, \bibinfo{author}{Hüttemann, M.}, \bibinfo{author}{Karimijafarbigloo, S.}, \bibinfo{author}{Wu, Y.}, \bibinfo{author}{Schmeink, A.}, \bibinfo{author}{Merhof, D.}, \bibinfo{year}{2023}.
\newblock \bibinfo{title}{Loss functions in the era of semantic segmentation: A survey and outlook}.
\newblock \href{http://arxiv.org/abs/2312.05391}{\tt arXiv:2312.05391}.
\bibitem[{Baena~J.(2015)}]{BJ2015WMDova}
\bibinfo{author}{Baena~J., e.a.}, \bibinfo{year}{2015}.
\newblock \bibinfo{title}{{Wm-dova maps for accurate polyp highlighting in colonoscopy: Validation vs. saliency maps from physicians}}.
\newblock \bibinfo{journal}{Computerized Medical Imaging and Graphics} , \bibinfo{pages}{99--111}.
\bibitem[{Dinh et~al.(2017)Dinh, Sohl-Dickstein and Bengio}]{dinh2017density}
\bibinfo{author}{Dinh, L.}, \bibinfo{author}{Sohl-Dickstein, J.}, \bibinfo{author}{Bengio, S.}, \bibinfo{year}{2017}.
\newblock \bibinfo{title}{Density estimation using real nvp}.
\newblock \href{http://arxiv.org/abs/1605.08803}{\tt arXiv:1605.08803}.
\bibitem[{Gal(2016)}]{Gal2016}
\bibinfo{author}{Gal, Y.}, \bibinfo{year}{2016}.
\newblock \bibinfo{title}{Uncertainty in deep learning}.
\newblock \bibinfo{journal}{University of Cambridge, Department of Engineering} .
\bibitem[{García-Farieta et~al.(2023)García-Farieta, Hortúa and Kitaura}]{garcíafarieta2023Bayesian}
\bibinfo{author}{García-Farieta, J.E.}, \bibinfo{author}{Hortúa, H.J.}, \bibinfo{author}{Kitaura, F.S.}, \bibinfo{year}{2023}.
\newblock \bibinfo{title}{Bayesian deep learning for cosmic volumes with modified gravity}.
\newblock \href{http://arxiv.org/abs/2309.00612}{\tt arXiv:2309.00612}.
\bibitem[{Globocan(2020)}]{WHO2020}
\bibinfo{author}{Globocan}, \bibinfo{year}{2020}.
\newblock \bibinfo{title}{Global health estimates 2020: Deaths by cause, age, sex, by country and by region, 2000-2019. who; 2020}.
\newblock \bibinfo{journal}{Accessed December, 2023} .
\bibitem[{Goodfellow et~al.(2016)Goodfellow, Bengio and Courville}]{goodfellow2016deep}
\bibinfo{author}{Goodfellow, I.}, \bibinfo{author}{Bengio, Y.}, \bibinfo{author}{Courville, A.}, \bibinfo{year}{2016}.
\newblock \bibinfo{title}{Deep Learning}.
\newblock Adaptive computation and machine learning, \bibinfo{publisher}{MIT Press}.
\newblock \URLprefix \url{https://books.google.co.in/books?id=Np9SDQAAQBAJ}.
\bibitem[{Guo et~al.(2017)Guo, Pleiss, Sun and Weinberger}]{pmlr-v70-guo17a}
\bibinfo{author}{Guo, C.}, \bibinfo{author}{Pleiss, G.}, \bibinfo{author}{Sun, Y.}, \bibinfo{author}{Weinberger, K.Q.}, \bibinfo{year}{2017}.
\newblock \bibinfo{title}{On calibration of modern neural networks}, in: \bibinfo{editor}{Precup, D.}, \bibinfo{editor}{Teh, Y.W.} (Eds.), \bibinfo{booktitle}{Proceedings of the 34th International Conference on Machine Learning}, \bibinfo{publisher}{PMLR}. pp. \bibinfo{pages}{1321--1330}.
\newblock \URLprefix \url{https://proceedings.mlr.press/v70/guo17a.html}.
\bibitem[{Hortúa et~al.(2020)Hortúa, Volpi, Marinelli and Malagò}]{Hortua2020}
\bibinfo{author}{Hortúa, H.J.}, \bibinfo{author}{Volpi, R.}, \bibinfo{author}{Marinelli, D.}, \bibinfo{author}{Malagò, L.}, \bibinfo{year}{2020}.
\newblock \bibinfo{title}{Parameter estimation for the cosmic microwave background with bayesian neural networks}.
\newblock \bibinfo{journal}{Physical Review D} \bibinfo{volume}{102}.
\newblock \URLprefix \url{http://dx.doi.org/10.1103/PhysRevD.102.103509}, \DOIprefix\doi{10.1103/physrevd.102.103509}.
\bibitem[{Iakubovskii(2019)}]{Yakubovskiy:2019}
\bibinfo{author}{Iakubovskii, P.}, \bibinfo{year}{2019}.
\newblock \bibinfo{title}{Segmentation models}.
\newblock \bibinfo{howpublished}{\url{https://github.com/qubvel/segmentation_models}}.
\bibitem[{Jadon(2020)}]{Jadon_2020}
\bibinfo{author}{Jadon, S.}, \bibinfo{year}{2020}.
\newblock \bibinfo{title}{A survey of loss functions for semantic segmentation}, in: \bibinfo{booktitle}{2020 IEEE Conference on Computational Intelligence in Bioinformatics and Computational Biology (CIBCB)}, \bibinfo{publisher}{IEEE}.
\newblock \URLprefix \url{http://dx.doi.org/10.1109/CIBCB48159.2020.9277638}, \DOIprefix\doi{10.1109/cibcb48159.2020.9277638}.
\bibitem[{Jha et~al.(2020)Jha, Riegler, Johansen, Halvorsen and Johansen}]{jha2020doubleunet}
\bibinfo{author}{Jha, D.}, \bibinfo{author}{Riegler, M.A.}, \bibinfo{author}{Johansen, D.}, \bibinfo{author}{Halvorsen, P.}, \bibinfo{author}{Johansen, H.D.}, \bibinfo{year}{2020}.
\newblock \bibinfo{title}{Doubleu-net: A deep convolutional neural network for medical image segmentation}.
\newblock \href{http://arxiv.org/abs/2006.04868}{\tt arXiv:2006.04868}.
\bibitem[{Jha et~al.(2021)Jha, Smedsrud, Johansen, de~Lange, Johansen, Halvorsen and Riegler}]{jha2021comprehensive}
\bibinfo{author}{Jha, D.}, \bibinfo{author}{Smedsrud, P.H.}, \bibinfo{author}{Johansen, D.}, \bibinfo{author}{de~Lange, T.}, \bibinfo{author}{Johansen, H.D.}, \bibinfo{author}{Halvorsen, P.}, \bibinfo{author}{Riegler, M.A.}, \bibinfo{year}{2021}.
\newblock \bibinfo{title}{A comprehensive study on colorectal polyp segmentation with resunet++, conditional random field and test-time augmentation}.
\newblock \href{http://arxiv.org/abs/2107.12435}{\tt arXiv:2107.12435}.
\bibitem[{Kwon et~al.(2020)Kwon, Won, Kim and Paik}]{KWON2020106816}
\bibinfo{author}{Kwon, Y.}, \bibinfo{author}{Won, J.H.}, \bibinfo{author}{Kim, B.J.}, \bibinfo{author}{Paik, M.C.}, \bibinfo{year}{2020}.
\newblock \bibinfo{title}{Uncertainty quantification using bayesian neural networks in classification: Application to biomedical image segmentation}.
\newblock \bibinfo{journal}{Computational Statistics and Data Analysis} \bibinfo{volume}{142}, \bibinfo{pages}{106816}.
\newblock \URLprefix \url{https://www.sciencedirect.com/science/article/pii/S016794731930163X}, \DOIprefix\doi{https://doi.org/10.1016/j.csda.2019.106816}.
\bibitem[{Lee et~al.(2017)Lee, Park, Kim, Lee, Sung, Song, Yoon and Moon}]{Lee2017RiskFactors}
\bibinfo{author}{Lee, J.}, \bibinfo{author}{Park, S.W.}, \bibinfo{author}{Kim, Y.S.}, \bibinfo{author}{Lee, K.J.}, \bibinfo{author}{Sung, H.}, \bibinfo{author}{Song, P.H.}, \bibinfo{author}{Yoon, W.J.}, \bibinfo{author}{Moon, J.S.}, \bibinfo{year}{2017}.
\newblock \bibinfo{title}{Risk factors of missed colorectal lesions after colonoscopy}.
\newblock \bibinfo{journal}{Medicine} \bibinfo{volume}{96}, \bibinfo{pages}{7468}.
\newblock \DOIprefix\doi{10.1097/MD.0000000000007468}.
\bibitem[{Li et~al.(2017)Li, Yang, Chen, Huang, Chen, Xu, Zhou, Zhong, Zhang and Wang}]{li_colorectal_2020}
\bibinfo{author}{Li, Q.}, \bibinfo{author}{Yang, G.}, \bibinfo{author}{Chen, Z.}, \bibinfo{author}{Huang, B.}, \bibinfo{author}{Chen, L.}, \bibinfo{author}{Xu, D.}, \bibinfo{author}{Zhou, X.}, \bibinfo{author}{Zhong, S.}, \bibinfo{author}{Zhang, H.}, \bibinfo{author}{Wang, T.}, \bibinfo{year}{2017}.
\newblock \bibinfo{title}{Colorectal polyp segmentation using a fully convolutional neural network}, in: \bibinfo{booktitle}{10th International Congress on Image and Signal Processing, BioMedical Engineering and Informatics (CISP-BMEI2017)}, \bibinfo{address}{Shangai, China}. pp. \bibinfo{pages}{14--17}.
\bibitem[{Li et~al.(2023)Li, Hu and Yang}]{li2023polypsam}
\bibinfo{author}{Li, Y.}, \bibinfo{author}{Hu, M.}, \bibinfo{author}{Yang, X.}, \bibinfo{year}{2023}.
\newblock \bibinfo{title}{Polyp-sam: Transfer sam for polyp segmentation}.
\newblock \href{http://arxiv.org/abs/2305.00293}{\tt arXiv:2305.00293}.
\bibitem[{Louizos and Welling(2017)}]{Louizos2017}
\bibinfo{author}{Louizos, C.}, \bibinfo{author}{Welling, M.}, \bibinfo{year}{2017}.
\newblock \bibinfo{title}{Multiplicative normalizing flows for variational bayesian neural networks}.
\newblock \bibinfo{journal}{34th international conference on machine learning} \bibinfo{volume}{70}, \bibinfo{pages}{2218--2227}.
\bibitem[{Ma et~al.(2021)Ma, Chen, Ng, Huang, Li, Li, Yang and Martel}]{LossOdyssey}
\bibinfo{author}{Ma, J.}, \bibinfo{author}{Chen, J.}, \bibinfo{author}{Ng, M.}, \bibinfo{author}{Huang, R.}, \bibinfo{author}{Li, Y.}, \bibinfo{author}{Li, C.}, \bibinfo{author}{Yang, X.}, \bibinfo{author}{Martel, A.L.}, \bibinfo{year}{2021}.
\newblock \bibinfo{title}{Loss odyssey in medical image segmentation}.
\newblock \bibinfo{journal}{Medical Image Analysis} \bibinfo{volume}{71}, \bibinfo{pages}{102035}.
\newblock \URLprefix \url{https://www.sciencedirect.com/science/article/pii/S1361841521000815}, \DOIprefix\doi{https://doi.org/10.1016/j.media.2021.102035}.
\bibitem[{Mei et~al.(2024)Mei, Zhou, Huang, Zhang, Zhou, Wu and Fu}]{mei2024survey}
\bibinfo{author}{Mei, J.}, \bibinfo{author}{Zhou, T.}, \bibinfo{author}{Huang, K.}, \bibinfo{author}{Zhang, Y.}, \bibinfo{author}{Zhou, Y.}, \bibinfo{author}{Wu, Y.}, \bibinfo{author}{Fu, H.}, \bibinfo{year}{2024}.
\newblock \bibinfo{title}{A survey on deep learning for polyp segmentation: Techniques, challenges and future trends}.
\newblock \href{http://arxiv.org/abs/2311.18373}{\tt arXiv:2311.18373}.
\bibitem[{Müller et~al.(2022)Müller, Soto-Rey and Kramer}]{müller2022guideline}
\bibinfo{author}{Müller, D.}, \bibinfo{author}{Soto-Rey, I.}, \bibinfo{author}{Kramer, F.}, \bibinfo{year}{2022}.
\newblock \bibinfo{title}{Towards a guideline for evaluation metrics in medical image segmentation} \href{http://arxiv.org/abs/2202.05273}{\tt arXiv:2202.05273}.
\bibitem[{NCI(2020)}]{seercolorectal}
\bibinfo{author}{NCI}, \bibinfo{year}{2020}.
\newblock \bibinfo{title}{Cancer stat facts: Colorectal cancer.}
\newblock \bibinfo{journal}{https://seer.cancer.gov/statfacts/html/colorect.html} \bibinfo{note}{Accessed December, 2023}.
\bibitem[{Simonyan et~al.(2014)Simonyan, Vedaldi and Zisserman}]{simonyan2014deep}
\bibinfo{author}{Simonyan, K.}, \bibinfo{author}{Vedaldi, A.}, \bibinfo{author}{Zisserman, A.}, \bibinfo{year}{2014}.
\newblock \bibinfo{title}{Deep inside convolutional networks: Visualising image classification models and saliency maps}.
\newblock \href{http://arxiv.org/abs/1312.6034}{\tt arXiv:1312.6034}.
\bibitem[{Tashk et~al.(2019)Tashk, Herp and Nadimi}]{Tashk2023AutomaticSegmentation}
\bibinfo{author}{Tashk, A.}, \bibinfo{author}{Herp, J.}, \bibinfo{author}{Nadimi, E.}, \bibinfo{year}{2019}.
\newblock \bibinfo{title}{Automatic segmentation of colorectal polyps based on a novel and innovative convolutional neural network approach}.
\newblock \bibinfo{journal}{WSEAS TRANSACTIONS on SYSTEMS and CONTROL} \bibinfo{volume}{12}.
\newblock \URLprefix \url{https://ml.sdu.dk/}.
\bibitem[{Wang et~al.(2022a)Wang, Gong and Wang}]{wang2022calibrating}
\bibinfo{author}{Wang, D.}, \bibinfo{author}{Gong, B.}, \bibinfo{author}{Wang, L.}, \bibinfo{year}{2022}a.
\newblock \bibinfo{title}{On calibrating semantic segmentation models: Analyses and an algorithm}.
\newblock \bibinfo{journal}{arXiv preprint arXiv:2212.12053} .
\bibitem[{Wang et~al.(2022b)Wang, Huang, Tang, Meng, Su and Song}]{wang2022stepwise}
\bibinfo{author}{Wang, J.}, \bibinfo{author}{Huang, Q.}, \bibinfo{author}{Tang, F.}, \bibinfo{author}{Meng, J.}, \bibinfo{author}{Su, J.}, \bibinfo{author}{Song, S.}, \bibinfo{year}{2022}b.
\newblock \bibinfo{title}{Stepwise feature fusion: Local guides global}.
\newblock \href{http://arxiv.org/abs/2203.03635}{\tt arXiv:2203.03635}.
\bibitem[{Wickstrøm et~al.(2020)Wickstrøm, Kampffmeyer and Jenssen}]{WICKSTROM2020101619}
\bibinfo{author}{Wickstrøm, K.}, \bibinfo{author}{Kampffmeyer, M.}, \bibinfo{author}{Jenssen, R.}, \bibinfo{year}{2020}.
\newblock \bibinfo{title}{Uncertainty and interpretability in convolutional neural networks for semantic segmentation of colorectal polyps}.
\newblock \bibinfo{journal}{Medical Image Analysis} \bibinfo{volume}{60}, \bibinfo{pages}{101619}.
\newblock \URLprefix \url{https://www.sciencedirect.com/science/article/pii/S1361841519301574}, \DOIprefix\doi{https://doi.org/10.1016/j.media.2019.101619}.
\bibitem[{Wu et~al.(2024)Wu, Lv, Chen, Hao and Li}]{wu2024colorectal}
\bibinfo{author}{Wu, Z.}, \bibinfo{author}{Lv, F.}, \bibinfo{author}{Chen, C.}, \bibinfo{author}{Hao, A.}, \bibinfo{author}{Li, S.}, \bibinfo{year}{2024}.
\newblock \bibinfo{title}{Colorectal polyp segmentation in the deep learning era: A comprehensive survey}.
\newblock \href{http://arxiv.org/abs/2401.11734}{\tt arXiv:2401.11734}.
\bibitem[{Xie et~al.(2021)Xie, Wang, Yu, Anandkumar, Alvarez and Luo}]{xie2021segformer}
\bibinfo{author}{Xie, E.}, \bibinfo{author}{Wang, W.}, \bibinfo{author}{Yu, Z.}, \bibinfo{author}{Anandkumar, A.}, \bibinfo{author}{Alvarez, J.M.}, \bibinfo{author}{Luo, P.}, \bibinfo{year}{2021}.
\newblock \bibinfo{title}{Segformer: Simple and efficient design for semantic segmentation with transformers}.
\newblock \href{http://arxiv.org/abs/2105.15203}{\tt arXiv:2105.15203}.
\bibitem[{Zou et~al.(2023)Zou, Chen, Yuan, Shen, Wang and Fu}]{zou2023review}
\bibinfo{author}{Zou, K.}, \bibinfo{author}{Chen, Z.}, \bibinfo{author}{Yuan, X.}, \bibinfo{author}{Shen, X.}, \bibinfo{author}{Wang, M.}, \bibinfo{author}{Fu, H.}, \bibinfo{year}{2023}.
\newblock \bibinfo{title}{A review of uncertainty estimation and its application in medical imaging}.
\newblock \href{http://arxiv.org/abs/2302.08119}{\tt arXiv:2302.08119}.

\end{thebibliography}

\appendix

\newpage

\section{Results: models with reparameterization trick}\label{appendix:Ap.B}

With respect to the qualitative importance of predictions in context of semantic segmentation in medical images, different comparative images of deterministic predictions, Bayesian predictions and uncertainty maps are shown, especially for cases where polyps are not easily detectable visually. The latter is used to show the advantage of implementing Bayesian neural networks in these cases. 
\vspace{2pt}
To compute masks for the BNN models, we take $n=50$ predictions over images and then average them. The resulting prediction is then binarized, with class 1 assigned if the prediction is greater than 0.5.

\begin{figure}[!ht]
\centering
\includegraphics[scale=.41]{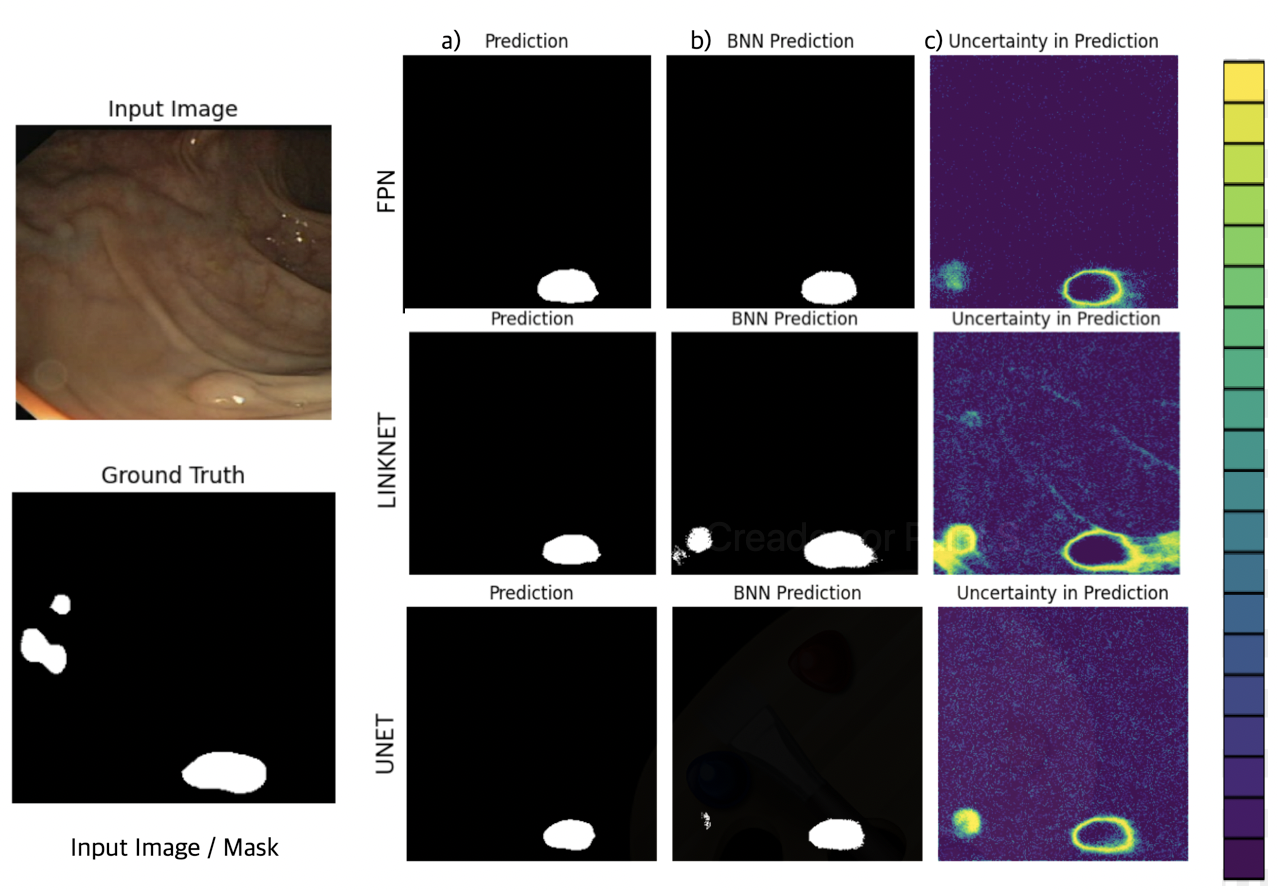}
\caption{(a) Deterministic prediction, (b) BNN prediction with reparameterization trick layers, and (c) Uncertainty maps for the same input image for UNET, FPN, and LINKNET architectures.
}
\label{fig:figura8}
\end{figure}

\section{Iteration results: deterministic models}\label{appendix:Ap.C}

A total of 36 iterations of deterministic models were run using Unet, Linknet, and FPN architectures. These models were tested with four different loss functions (total loss, binary cross-entropy, Jaccard loss, and Dice loss) and implemented with three different backbones for each possible combination. By incorporating these variants of loss functions into training process, the aim was to guarantee a comprehensive evaluation and enhance the adaptability of the models.
From results obtained, we saw that EfficientNetB7 was best backbone in terms of performance. In particular, loss function \textit{Binary cross-entropy} proved to be the most effective for Unet and Linknet architectures, while for FPN, the loss function  \textit{total loss} stood out as the best choice.

\begin{table}[!h]
\caption{\label{tab3}Results in test dataset for deterministic Unet iterations}
\vspace{4pt}
  \scalebox{0.77}{ \begin{tabular}{ccllll}
\hline
\begin{tabular}[c]{@{}c@{}}Model  UNET\\ +\\ Backbones\end{tabular} &
  Loss functions &
  IOU &
  Recall &
  \begin{tabular}[c]{@{}l@{}}False \\ negatives\end{tabular} &
  \begin{tabular}[c]{@{}l@{}}False \\ positives\end{tabular} \\ \hline
EfficientNetB7 &
  \begin{tabular}[c]{@{}c@{}}Dice loss\\ \\ Jaccard loss\\ \\ Total loss\\ \\ Binary cross-entropy \vspace{3pt}\end{tabular} &
  \begin{tabular}[c]{@{}l@{}}0.918\\ \\ 0.916\\ \\ 0.914\\ \\ \textbf{0.933}\end{tabular} &
  \begin{tabular}[c]{@{}l@{}}0.903\\ \\ 0.910\\ \\ 0.89\\ \\ \textbf{0.911}\end{tabular} &
  \begin{tabular}[c]{@{}l@{}}65794\\ \\ 60686\\ \\ 74400\\ \\ 62053\end{tabular} &
  \begin{tabular}[c]{@{}l@{}}33849\\ \\ 42789\\ \\ 29512\\ \\ 17903\end{tabular} \\ \hline
SeresNet101 &
  \begin{tabular}[c]{@{}c@{}}Dice loss\\ \\ Jaccard loss\\ \\ Total loss\\ \\ Binary cross-entropy \vspace{3pt}\end{tabular} &
  \begin{tabular}[c]{@{}l@{}}0.874\\ \\ 0.858\\ \\ 0.856\\ \\ 0.868\end{tabular} &
  \begin{tabular}[c]{@{}l@{}}0.832\\ \\ 0.799\\ \\ 0.764\\ \\ 0.89\end{tabular} &
  \begin{tabular}[c]{@{}l@{}}114020\\ \\ 136452\\ \\ 159909\\ \\ 74786\end{tabular} &
  \begin{tabular}[c]{@{}l@{}}41629\\ \\ 35358\\ \\ 11227\\ \\ 99174\end{tabular} \\ \hline
DenseNet169 &
  \begin{tabular}[c]{@{}c@{}}Dice loss\\ \\ Jaccard loss\\ \\ Total loss\\ \\ Binary cross-entropy \vspace{3pt}\end{tabular} &
  \begin{tabular}[c]{@{}l@{}}0.876\\ \\ 0.832\\ \\ 0.813\\ \\ 0.869\end{tabular} &
  \begin{tabular}[c]{@{}l@{}}0.883\\ \\ 0.736\\ \\ 0.696\\ \\ 0.795\end{tabular} &
  \begin{tabular}[c]{@{}l@{}}79532\\ \\ 179126\\ \\ 205932\\ \\ 139299\end{tabular} &
  \begin{tabular}[c]{@{}l@{}}80046\\ \\ 24109\\ \\ 19740\\ \\ 18133\end{tabular} \\ \hline
\end{tabular}}
\end{table}

\begin{table}[!t]
\caption{\label{tab4}Results in test dataset for deterministic FPN iterations}
\vspace{4pt}
  \scalebox{0.8}{
\begin{tabular}{ccllll}
\hline
\begin{tabular}[c]{@{}c@{}}Model  FPN\\ +\\ Backbones\end{tabular} &
  Loss functions &
  IOU &
  Recall &
  \begin{tabular}[c]{@{}l@{}}False \\ negatives\end{tabular} &
  \begin{tabular}[c]{@{}l@{}}False \\ positives\end{tabular} \\ \hline
EfficientNetB7 &
  \begin{tabular}[c]{@{}c@{}}Dice loss\\ \\ Jaccard loss\\ \\ Total loss\\ \\ Binary cross-entropy \vspace{3pt}\end{tabular} &
  \begin{tabular}[c]{@{}l@{}}0.910\\ \\ 0.920\\ \\ \textbf{0.930}\\ \\ 0.891\end{tabular} &
  \begin{tabular}[c]{@{}l@{}}0.888\\ \\ 0.900\\ \\ \textbf{0.941}\\ \\ 0.860\end{tabular} &
  \begin{tabular}[c]{@{}l@{}}75794\\ \\ 63534\\ \\ 43275\\ \\ 94798\end{tabular} &
  \begin{tabular}[c]{@{}l@{}}29173\\ \\ 31787\\ \\ 36810\\ \\ 17903\end{tabular} \\ \hline
SeresNet101 &
  \begin{tabular}[c]{@{}c@{}}Dice loss\\ \\ Jaccard loss\\ \\ Total loss\\ \\ Binary cross-entropy \vspace{3pt}\end{tabular} &
  \begin{tabular}[c]{@{}l@{}}0.890\\ \\ 0.823\\ \\ 0.890\\ \\ 0.901\end{tabular} &
  \begin{tabular}[c]{@{}l@{}}0.860\\ \\ 0.714\\ \\ 0.861\\ \\ 0.880\end{tabular} &
  \begin{tabular}[c]{@{}l@{}}94798\\ \\ 193914\\ \\ 94623\\ \\ 75741\end{tabular} &
  \begin{tabular}[c]{@{}l@{}}35999\\ \\ 18960\\ \\ 31155\\ \\ 38149\end{tabular} \\ \hline
DenseNet169 &
  \begin{tabular}[c]{@{}c@{}}Dice loss\\ \\ Jaccard loss\\ \\ Total loss\\ \\ Binary cross-entropy \vspace{3pt}\end{tabular} &
  \begin{tabular}[c]{@{}l@{}}0.801\\ \\ 0.820\\ \\ 0.780\\ \\ 0.920\end{tabular} &
  \begin{tabular}[c]{@{}l@{}}0.702\\ \\ 0.740\\ \\ 0.720\\ \\ 0.920\end{tabular} &
  \begin{tabular}[c]{@{}l@{}}201255\\ \\ 171655\\ \\ 273073\\ \\ 53301\end{tabular} &
  \begin{tabular}[c]{@{}l@{}}41515\\ \\ 54521\\ \\ 175285\\ \\ 38920\end{tabular} \\ \hline
\end{tabular}}
\end{table}

\begin{table}[!ht]
\caption{\label{tab5}Results in test dataset for deterministic Linknet iterations}
\vspace{4pt}
  \scalebox{0.8}{
\begin{tabular}{ccllll}
\hline
\begin{tabular}[c]{@{}c@{}}Model  Linknet\\ +\\ Backbones\end{tabular} &
  Loss functions &
  IOU &
  Recall &
  \begin{tabular}[c]{@{}l@{}}False \\ negatives\end{tabular} &
  \begin{tabular}[c]{@{}l@{}}False \\ positives\end{tabular} \\ \hline
EfficientNetB7 &
  \begin{tabular}[c]{@{}c@{}}Dice loss\\ \\ Jaccard loss\\ \\ Total loss\\ \\ Binary cross-entropy \vspace{3pt}\end{tabular} &
  \begin{tabular}[c]{@{}l@{}}0.937\\ \\ 0.920\\ \\ 0.915\\ \\ \textbf{0.941}\end{tabular} &
  \begin{tabular}[c]{@{}l@{}}0.925\\ \\ 0.91\\ \\ 0.881\\ \\ \textbf{0.927}\end{tabular} &
  \begin{tabular}[c]{@{}l@{}}50752\\ \\ 55041\\ \\ 80798\\ \\ 49557\end{tabular} &
  \begin{tabular}[c]{@{}l@{}}25235\\ \\ 30667\\ \\ 21907\\ \\ 20749\end{tabular} \\ \hline
SeresNet101 &
  \begin{tabular}[c]{@{}c@{}}Dice loss\\ \\ Jaccard loss\\ \\ Total loss\\ \\ Binary cross-entropy \vspace{3pt}\end{tabular} &
  \begin{tabular}[c]{@{}l@{}}0.840\\ \\ 0.836\\ \\ 0.873\\ \\ 0.855\end{tabular} &
  \begin{tabular}[c]{@{}l@{}}0.760\\ \\ 0.746\\ \\ 0.829\\ \\ 0.760\end{tabular} &
  \begin{tabular}[c]{@{}l@{}}160425\\ \\ 171996\\ \\ 115780\\ \\ 161860\end{tabular} &
  \begin{tabular}[c]{@{}l@{}}33466\\ \\ 26434\\ \\ 40529\\ \\ 10679\end{tabular} \\ \hline
DenseNet169 &
  \begin{tabular}[c]{@{}c@{}}Dice loss\\ \\ Jaccard loss\\ \\ Total loss\\ \\ Binary cross-entropy \vspace{3pt}\end{tabular} &
  \begin{tabular}[c]{@{}l@{}}0.789\\ \\ 0.863\\ \\ 0.813\\ \\ 0.897\end{tabular} &
  \begin{tabular}[c]{@{}l@{}}0.80\\ \\ 0.81\\ \\ 0.697\\ \\ 0.87\end{tabular} &
  \begin{tabular}[c]{@{}l@{}}133360\\ \\ 123006\\ \\ 205700\\ \\ 87808\end{tabular} &
  \begin{tabular}[c]{@{}l@{}}163537\\ \\ 47313\\ \\ 18456\\ \\ 38144\end{tabular} \\ \hline
\end{tabular}}
\end{table}

\end{document}